\documentclass[12pt,a4paper]{article}
\usepackage{axodraw}
\usepackage{epsfig}

\let\log\ln

\advance\textheight by \topskip
\if@twoside \oddsidemargin -17pt \evensidemargin 00pt
\else \oddsidemargin 00pt \evensidemargin 00pt
\fi
\expandafter\ifx\csname mathrm\endcsname\relax\def\mathrm#1{{\rm #1}}\fi

\renewcommand{\theequation}{\thesection.\arabic{equation}}
\newcounter{saveeqn}

\makeatletter 
\@addtoreset{equation}{section}
\makeatother

\def\beq{\begin{equation}}
\def\eeq{\end{equation}}
\def\beqar{\begin{eqnarray}}
\def\eeqar{\end{eqnarray}}
\def\barr#1{\begin{array}{#1}}
\def\earr{\end{array}}
\def\bfi{\begin{figure}}
\def\efi{\end{figure}}
\def\btab{\begin{table}}
\def\etab{\end{table}}
\def\bce{\begin{center}}
\def\ece{\end{center}}

\def\nl{\nonumber\\}

\def\si{\sigma}

\def\reffi#1{\mbox{Fig.~\ref{#1}}}

\def\refse#1{\mbox{Sect.~\ref{#1}}}

\def\refapp#1{\mbox{Appendix~\ref{#1}}}
\def\citere#1{\mbox{Ref.~\cite{#1}}}
\def\citeres#1{\mbox{Refs.~\cite{#1}}}

\def\solid{\raise.9mm\hbox{\protect\rule{1.1cm}{.2mm}}}
\def\dash{\raise.9mm\hbox{\protect\rule{2mm}{.2mm}}\hspace*{1mm}}
\def\dot{\rlap{$\cdot$}\hspace*{2mm}}

\def\solid{\raise.9mm\hbox{\protect\rule{12mm}{.2mm}}}
\def\dash{\raise.9mm\hbox{\protect\rule{1.6mm}{.2mm}}\hspace*{1mm}}
\def\dot{\raise.9mm\hbox{\protect\rule{0.8mm}{.2mm}}\hspace*{0.8mm}}
\def\dashdot{\raise.9mm\hbox{\protect\rule{.3mm}{.2mm}}\hspace*{.8mm}\raise.9mm\hbox{\protect\rule{1.3mm}{.2mm}}\hspace*{.8mm}}

\newcommand{\GeV}{\unskip\,\mathrm{GeV}}

\newcommand{\TeV}{\unskip\,\mathrm{TeV}}
\newcommand{\fba}{\unskip\,\mathrm{fb}}

\def\mathswitchr#1{\relax\ifmmode{\mathrm{#1}}\else$\mathrm{#1}$\fi}
\newcommand{\PM}{\mathswitchr M}
\newcommand{\Pm}{\mathswitchr m}
\newcommand{\PB}{\mathswitchr B}
\newcommand{\PV}{\mathswitch V}

\newcommand{\PW}{\mathswitchr W}
\newcommand{\Pw}{\mathswitchr w}
\newcommand{\Pz}{\mathswitchr z}
\newcommand{\PZ}{\mathswitchr Z}

\newcommand{\Pg}{\mathswitchr g}
\newcommand{\PH}{\mathswitchr H}
\newcommand{\Pe}{\mathswitchr e}

\newcommand{\Pd}{\mathswitchr d}
\newcommand{\PD}{\mathswitchr D}

\newcommand{\Pu}{\mathswitchr u}
\newcommand{\PU}{\mathswitchr U}
\newcommand{\Ps}{\mathswitchr s}
\newcommand{\Pb}{\mathswitchr b}
\newcommand{\Pc}{\mathswitchr c}
\newcommand{\Pt}{\mathswitchr t}

\newcommand{\Pp}{\mathswitchr {p}}

\def\mathswitch#1{\relax\ifmmode#1\else$#1$\fi}

\newcommand{\MW}{\mathswitch {M_\PW}}

\newcommand{\MZ}{\mathswitch {M_\PZ}}

\newcommand{\Mb}{\mathswitch {m_\Pb}}
\newcommand{\Mt}{\mathswitch {m_\Pt}}
\newcommand{\GW}{\mathswitch {\Gamma_\PW}}
\newcommand{\GZ}{\Gamma_{\PZ}}
\newcommand{\PL}{\mathswitch {P_\PL}}

\newcommand{\scrs}{\scriptscriptstyle}
\newcommand{\sw}{\mathswitch {s_{\scrs\PW}}}
\newcommand{\cw}{\mathswitch {c_{\scrs\PW}}}

\def\ie{i.e.\ }

\newcommand{\rd}{{\mathrm{d}}}

\renewcommand{\L}{{\cal L}}

\newcommand{\CM}{\mathrm{CM}}

\newcommand{\SM}{{\mathrm{SM}}}

\newcommand{\PT}{P_{\mathrm{T}}}
\newcommand{\ET}{E_{\mathrm{T}}}
\newcommand{\PTmiss}{P_{\mathrm{T}}^{\mathrm{miss}}}

\newcommand{\PTcut}{P_{\mathrm{T}}^{\mathrm{cut}}}
\newcommand{\MT}{M_{\mathrm{T}}}

\makeatletter
\newcount\@tempcntc
\def\@citex[#1]#2{\if@filesw\immediate\write\@auxout{\string\citation{#2}}\fi
  \@tempcnta\z@\@tempcntb\m@ne\def\@citea{}\@cite{\@for\@citeb:=#2\do
    {\@ifundefined
       {b@\@citeb}{\@citeo\@tempcntb\m@ne\@citea
        \def\@citea{,\penalty\@m\ }{\bf ?}\@warning
       {Citation `\@citeb' on page \thepage \space undefined}}%
    {\setbox\z@\hbox{\global\@tempcntc0\csname
b@\@citeb\endcsname\relax}%
     \ifnum\@tempcntc=\z@ \@citeo\@tempcntb\m@ne
       \@citea\def\@citea{,\penalty\@m}
       \hbox{\csname b@\@citeb\endcsname}%
     \else
      \advance\@tempcntb\@ne
      \ifnum\@tempcntb=\@tempcntc
      \else\advance\@tempcntb\m@ne\@citeo
      \@tempcnta\@tempcntc\@tempcntb\@tempcntc\fi\fi}}\@citeo}{#1}}

\def\@citeo{\ifnum\@tempcnta>\@tempcntb\else\@citea
  \def\@citea{,\penalty\@m}%
  \ifnum\@tempcnta=\@tempcntb\the\@tempcnta\else
   {\advance\@tempcnta\@ne\ifnum\@tempcnta=\@tempcntb \else
\def\@citea{--}\fi
    \advance\@tempcnta\m@ne\the\@tempcnta\@citea\the\@tempcntb}\fi\fi}
\makeatother

\def\draftdate{\relax}
\def\mpar#1{\relax}
\def\mda{\relax}
\def\mua{\relax}
\def\mla{\relax}
\def\draft{
\def\thtystars{******************************}
\def\sixtystars{\thtystars\thtystars}
\typeout{}
\typeout{\sixtystars**}
\typeout{* Draft mode!
         For final version remove \protect\draft\space in source file *}
\typeout{\sixtystars**}
\typeout{}
\def\draftdate{\today}
\def\mua{\marginpar[\boldmath\hfil$\uparrow$]%
                   {\boldmath$\uparrow$\hfil}%
                    \typeout{marginpar: $\uparrow$}\ignorespaces}
\def\mda{\marginpar[\boldmath\hfil$\downarrow$]%
                   {\boldmath$\downarrow$\hfil}%
                    \typeout{marginpar: $\downarrow$}\ignorespaces}
\def\mla{\marginpar[\boldmath\hfil$\rightarrow$]%
                   {\boldmath$\leftarrow $\hfil}%
                    \typeout{marginpar: $\leftrightarrow$}\ignorespaces}
\def\Mua{\marginpar[\boldmath\hfil$\Uparrow$]%
                   {\boldmath$\Uparrow$\hfil}%
                    \typeout{marginpar: $\Uparrow$}\ignorespaces}
\def\Mda{\marginpar[\boldmath\hfil$\Downarrow$]%
                   {\boldmath$\Downarrow$\hfil}%
                    \typeout{marginpar: $\Downarrow$}\ignorespaces}
\def\Mla{\marginpar[\boldmath\hfil$\Rightarrow$]%
                   {\boldmath$\Leftarrow $\hfil}%
                    \typeout{marginpar: $\Leftrightarrow$}\ignorespaces}
\def\mpar##1{\marginpar{\hbadness10000%
                      \sloppy\hfuzz10pt\boldmath\bf##1}%
                      \typeout{marginpar: ##1}\ignorespaces}
\overfullrule 5pt
\oddsidemargin -15mm
\marginparwidth 29mm
}

\newcommand{\thismonth}{\ifcase\month\or January\or February\or March \or April
\or May \or June \or July \or August \or September \or \November \or 
\December\fi}
\makeatletter

\def\eqnarray{\stepcounter{equation}\let\@currentlabel=\theequation
\global\@eqnswtrue
\global\@eqcnt\z@\tabskip\@centering\let\\=\@eqncr
$$\halign to \displaywidth\bgroup\hskip\@centering
  $\displaystyle\tabskip\z@{##}$\@eqnsel&\global\@eqcnt\@ne
  \hskip 2\arraycolsep \hfil${##}$\hfil
  &\global\@eqcnt\tw@ \hskip 2\arraycolsep $\displaystyle\tabskip\z@{##}$\hfil
   \tabskip\@centering&\llap{##}\tabskip\z@\cr}
\def\appendix{\par
 \setcounter{section}{0} \setcounter{subsection}{0}
 \def\thesection{\Alph{section}}}

\makeatother

\def\eqalign#1{\null\,\vcenter{\openup\jot\ialign{\strut\hfil$%
        \displaystyle{##{}}$&$\displaystyle{{}##}$\hfil\crcr#1\crcr}}\,}

\def\slash#1{\setbox0\hbox{$#1$}\hbox to\wd0{\hss$/$\hss}\nobreak\hskip-\wd0\box0}
\def\fmruletab#1{{\openup1ex\halign{\hskip 1cm\hfil$\vcenter{\hsize=50pt\noindent##}$\hfil &\hskip 1cm $\displaystyle##$\hfil\cr#1}}}
\def\textscr#1{\textrm{\scriptsize #1}}
\def\tfrac#1#2{\textstyle\frac{#1}{#2}}

\begin{document}

\tolerance=100000
\thispagestyle{empty}
\setcounter{page}{0}

\thispagestyle{empty}
\def\thefootnote{\fnsymbol{footnote}}
\setcounter{footnote}{1}
\null
\draftdate\hfill  DFTT 06/2006
\\
%\strut\hfill ZU-TH 19/05 \\
%\strut\hfill DFTT 36/05\\
\strut\hfill hep-ph/0604273
\vskip 0cm
\vfill
\begin{center}
%  {\Large \bf A Quasi-Gauge-Invariant signal definition
  {\Large \bf Pseudo-observables in Axial gauge
\par} \vskip 2.5em
{\large
{\sc E. Accomando}}%
\\[.5cm]
{\it Dipartimento di Fisica Teorica, Universit\`a di Torino,\\
and INFN, Sezione di Torino,\\
Via P. Giuria 1, 10125 Torino, Italy}
\\[0.3cm]
\par
\end{center}\par
\vskip 2.0cm \vfill {\bf Abstract:} \par 

We have given a first application of the Axial gauge \`a la Dams and Kleiss 
to the Standard Model ($\SM$) physics at the LHC. We have focused on the issue 
of providing a 
well-behaved signal definition in presence of potentially strong gauge 
cancellations at high energies. As a first illustration, we have analysed the 
production of 
$\PW\PZ$ vector-boson pairs, which gives rise to four final-state fermions.
Purely leptonic final states, $\Pp\Pp\to l\bar\nu_ll^\prime\bar{l^\prime}$, 
have been numerically investigated in the region of high center-of-mass 
energies and large scattering angles, particularly sensitive to gauge 
dependences. We have found that the Axial gauge is the appropriate framework 
to recover a meaningful separation of signal and irreducible background over
the full energy domain.
\par
\vskip 1cm
\noindent
April 2006
\par
\null
\setcounter{page}{0}
\clearpage
\def\thefootnote{\arabic{footnote}}
\setcounter{footnote}{0}

\def\mla{\marginpar[\boldmath\hfil$\rightarrow$]%
                   {\boldmath$\leftarrow $\hfil}%
                    \typeout{marginpar: $\leftrightarrow$}\ignorespaces}
\def\mua{\marginpar[\boldmath\hfil$\uparrow$]%
                   {\boldmath$\uparrow$\hfil}%
                    \typeout{marginpar: $\uparrow$}\ignorespaces}
\def\mda{\marginpar[\boldmath\hfil$\downarrow$]%
                   {\boldmath$\downarrow$\hfil}%
                    \typeout{marginpar: $\downarrow$}\ignorespaces}

\section{Introduction}
\label{Introduction}

This letter deals with the phenomenology of the $\SM$ electroweak interactions 
at high energy scales. 
%Despite the role of the parton distribution functions (PDF) in decreasing the 
%statistics, the 
The high energy region has an enormous potential for particle discovery. A 
large set of new 
signatures is expected in this kinematical domain at the upcoming and future 
colliders. Of course, the signals might be very complicated and the background
overwhelming, expecially in hadronic environments. 

With increasing the energy, new channels with many particles in the final 
state will indeed open up, making difficult to understand the underlying 
physics. In this intricate context, the comparison between 
measurements and theoretical predictions will be far from easy task. A long 
chain of Monte Carlo simulations will be employed to deconvolute the observed 
quantities back to the partonic variables. With this prospect, identifying 
the signal
configuration and picking out the kinematical regions where it is expected to 
be enhanced over the background could probe decisive in the data analysis.

In this letter, the question we want to address is precisely how to disentagle 
the signal from its irreducible background. Commonly, what we consider as a 
signal is represented
by a subset of Feynman diagrams which describes the particles we are searching 
for as intermediate states. In most of the cases, this sub-contribution 
is not separately gauge invariant. The signal may indeed contain 
gauge-invariance-breaking terms which are only cancelled against their 
irreducible-background counterpart in the total amplitude. 

In principle, any bare selection of signal is not theoretically well-defined;
only gauge-independent quantities can be related to physical observables.
However, questions of principles are often of scarce practical relevance.
The point is to evaluate the numerical impact of the potentially 
badly-behaving terms. The answer is influenced by different factors. It varies 
according to the process at hand, the energy scale the reaction occurs at, and 
the gauge-fixing choice.  

It is quite a known fact that at LEP2 energies the gauge-invariance-breaking 
terms are generally unimportant, when computed in the 't Hooft Feynman gauge.
But, they might cause strong gauge cancellations between the various Feynman 
diagrams contributing to a given process at higher energy scales. This 
phenomenon is more and more enhanced as the off-shellness of the 
intermediate-state particles and the number of graphs increase 
\cite{kleiss_stirling,adp,phase}.
Complex processes with many particles in the final state might thus undergo 
huge interferences, making it senseless any signal selection.

In this letter we show that considering the Standard Model in the Axial gauge 
\`a la Dams and Kleiss \cite{dk} allows one to recover a quasi-gauge-invariant 
signal definition. In order to discuss this issue, we focus on the production 
of $\PW\PZ$ gauge-boson pairs with large invariant mass $\PM_{\PW\PZ}$ at the 
upcoming Large Hadron Collider (LHC), giving rise to four-fermion final 
states. The signal definition for this kind of processes has a well 
established reference. It has been in fact stated and largely used at LEP2 for 
the analysis of $\PW\PW$ and $\PZ\PZ$ physics.

The interest in the $\PW\PZ$ process is not only in giving a typical example
of high energy electroweak phenomenology. The LHC will in fact collect 
thousands of di-boson events \cite{Haywood:1999qg}, hence giving prospects for 
a detailed investigation of the $\PW\PW\PZ$ trilinear couplings in this 
channel. Possible anomalous self-interactions, which parametrize deviations 
from $\SM$ predictions due to new physics occurring at $\TeV$ scales, are 
indeed expected to increasingly enhance the gauge-boson pair-production cross 
section at large di-boson invariant masses. Extracting the signal is thus of 
vital importance to measure the involved trilinear gauge coupling. 

The paper is organized as follows: in \refse{sec:axialgauge} we briefly 
describe the axial gauge. The general setup of our numerical analysis
is given in \refse{sec:processes}. In \refse{sec:signal}, we discuss the 
possibility of a well-behaved signal definition, comparing the results in 
Unitary and Axial gauge. Our findings are summarized in 
\refse{sec:conclusions}. The SM Feynman rules in the axial gauge are listed in 
\refapp{sec:feynmanrules}.

\section{Axial gauge}
\label{sec:axialgauge}

One of the most appealing reasons for computing SM processes in the Axial gauge
is that it can provide a more severe check on gauge invariance (see for 
instance \citere{gauge_check}). In the following sections, we point out a 
further advantage, namely the possibility to minimize the gauge cancellations 
between Feynman diagrams at high energies. Here, we simply give a 
brief description of the Axial gauge content.

The formalism is not exceedingly cumbersome. There are indeed unphysical 
bosonic particles, as intermediate states, but no Fadeev-Popov ghosts.   
Moreover, two realizations are possible. The first one keeps the bilinear terms
in the unphysical bosons and the $\PW$ or $\PZ$ particles, giving rise to
mixed propagators \cite{kunszt-soper}. The latter has diagonalized 
propagators, but new interaction vertices \cite{dk}. In the following, we
discuss and use this latter approach.

The Axial gauge manifests its nature in the gauge-fixing part of the 
lagrangian
\beq
\L_{\textscr{gauge-fixing}}=
	-\tfrac12\lambda n^\mu A^a_\mu A^a_\nu n^\nu
   -\tfrac12\lambda (n\cdot B)^2,
\eeq
where $A^a_\mu$ ($a$=1,2,3) are the SU(2) gauge fields, and $B_\mu$ belongs
to U(1). The four-vector $n_\mu$ represents the gauge invariance control 
parameter, the physical observables must be independent of. The resulting 
Feynman rules, obtained in the limit~$\lambda\to\infty$, are summarized in 
\refapp{sec:feynmanrules}. 

Once rewritten $A^3$ and $B$ in terms of the physical fields $\PZ$ and 
$\gamma$, and parametrizing the Higgs-doublet field as
\beq
\phi =\frac1{\sqrt 2}\pmatrix{\sqrt 2\phi_\PW\cr {v+\PH+i\phi_\PZ}\cr},~~~~~~~v=2\sqrt{-\mu^2/\lambda_\phi}
\eeq
where $\PH$ represents the Higgs field with mass $\PM_\PH$ and $\lambda_\phi$ 
the Higgs self-coupling, the mixing terms between physical and unphysical 
neutral fields are
\beq
\eqalign{
\L_{\PZ\phi_\PZ,\textscr{bilinear}}&=
		-\tfrac12(\partial^\nu\PZ^\mu)(\partial_\nu\PZ_\mu)
      +\tfrac12(\partial^\mu\PZ_\mu)(\partial^\nu\PZ_\nu)
		+\tfrac12\PM_\PZ^2\PZ_\mu\PZ^\mu\cr
		&\qquad-\tfrac12\lambda n^\mu\PZ_\mu\PZ_\nu n^\nu
		+\tfrac12(\partial^\mu\phi_\PZ)(\partial_\mu\phi_\PZ)
		-\PM_\PZ\PZ^\mu\partial_\mu\phi_\PZ.
}
\eeq
This part of the Lagrangian can be diagonalized in momentum space by
applying the following transformation
\beq
\phi_\PZ(k)\to\phi_\PZ(k)+2i\PM_\PZ\frac{k^\mu \PZ_\mu(k)}{k^2}.
\eeq
After the diagonalization, the quadratic terms in the Lagrangian
for the field~$\PZ$ give rise to the $\PZ$-boson propagator
\beq
\Delta_{\nu\mu}=\frac{-i\left(
      g_{\nu\mu}
      -\frac{n_\nu k_\mu+n_\mu k_\nu}{n\cdot k}
      +k_\nu k_\mu\frac{n^2+(k^2-\PM_\PZ^2)/\lambda}{(n\cdot k)^2}
   \right)}{k^2-\PM_\PZ^2+i\epsilon}.
\eeq
Taking the limit $\lambda\to\infty$, one recovers the expression reported
in \refapp{sec:propagators}. An analogous procedure applies to the 
$\PW$-boson, and gives back the same propagator with $\PM_\PZ$ replaced by 
$\PM_\PW$.

The boson propagators in axial gauge display the peculiar property of being 
well-behaved at high energy. In the unitary gauge, the term 
$k_\mu k_\nu/\PM_\PV^2$ ($\PV =\PW,\PZ$) appearing in the numerator of the 
propagator leads to gauge-invariance-breaking terms of order $s/\PM_\PV^2$ 
(with $s$ the center-of-mass energy squared) in individual Feynman diagrams. 
By contrast, in axial gauge
each numerator factor $k_\mu$ is suppressed by a corresponding factor 
$1/(k\cdot n)$, preventing the growth with energy of individual diagrams and
the subsequent appearence of strong gauge cancellations between them.  
This important property, shared also by the new vertices, and its 
phenomenological consequences are the focus of this letter.

\section{Setup of the numerical analysis}
\label{sec:processes}

We consider the class of processes 
$\Pp\Pp\to l\bar\nu_ll^\prime\bar{l^\prime}$, where $l,l^\prime=\Pe$ or $\mu$. 
In our notation, $l\bar\nu_l$ indicates both $l^-\bar\nu_l$ and $l^+\nu_l$. 
These processes are characterized by three isolated charged leptons plus 
missing energy in the final state. They include $\PW\PZ$ production as 
intermediate state.  

Since the two incoming hadrons are protons and we sum over final states with 
opposite charges, we find
\beqar\label{eq:convol}%\refeq{eq:convol}
\rd\si^{h_1h_2}(P_1,P_2,p_f) = 
\int_0^1\rd x_1 \rd x_2 &&\sum_{U=u,c}\sum_{D=d,s}
\Bigl[f_{\bar\PD,\Pp}(x_1,Q^2)f_{\PU,\Pp}(x_2,Q^2)\,\rd\hat\si^{\bar\PD\PU}
(x_1P_1,x_2P_2,p_f)
\nl&&{}
+f_{\bar\PU,\Pp}(x_1,Q^2)f_{\PD,\Pp}(x_2,Q^2)\,\rd\hat\si^{\bar\PU\PD}
(x_1P_1,x_2P_2,p_f)
\nl&&{}
+f_{\bar\PD,\Pp}(x_2,Q^2)f_{\PU,\Pp}(x_1,Q^2)\,\rd\hat\si^{\bar\PD\PU}
(x_2P_2,x_1P_1,p_f)
\nl&&{}
+f_{\bar\PU,\Pp}(x_2,Q^2)f_{\PD,\Pp}(x_1,Q^2)\,\rd\hat\si^{\bar\PU\PD}
(x_2P_2,x_1P_1,p_f)
\Bigr]
\eeqar
in leading order of QCD.

For the masses we use the input values \cite{Hagiwara:pw}:
%\beq
%\begin{array}[b]{lcllcllcl}
%\MW & = & 80.425\GeV, \qquad &
%\MZ & = & 91.1876\GeV, \\
%\Mt & = & 178.0\GeV, &
%\Mb & = & 4.9\GeV.
%\end{array}
%\label{eq:SMpar}
%\eeq
\beq
\MW = 80.425\GeV,~~~~\MZ = 91.1876\GeV,~~~~\Mb = 4.9\GeV.
\label{eq:SMpar}
\eeq
All fermions but the b-quark are taken to be massless.

The weak mixing angle is fixed by $\sw^2=1-\MW^2/\MZ^2$.  Moreover, we
adopted the so called $G_{\mu}$-scheme, which effectively includes
higher-order contributions associated with the running of the
electromagnetic coupling and the leading universal two-loop
$\Mt$-dependent corrections. This corresponds to parametrize the
lowest-order matrix element in terms of the effective coupling
$\alpha_{G_{\mu}}=\sqrt{2}G_{\mu}\MW^2\sw^2/\pi$.

Additional input parameters are the quark-mixing matrix elements whose
values have been taken to be $|V_{\Pu\Pd}|=0.974$ \cite{Hocker:2001xe}, 
$|V_{\Pc\Ps}|=|V_{\Pu\Pd}|$, $|V_{\Pu\Ps}|=|V_{\Pc\Pd}|=
\sqrt{1- |V_{\Pu\Pd}|^2}$, and zero for all other relevant matrix elements.

For the numerical results presented here, we have used the fixed-width scheme 
with $\GZ$ and $\GW$ from standard formulas
\beqar
\GZ &=& {\alpha\MZ\over{24\sw^2\cw^2}} 
\Bigl[21-40\sw^2+{160\over 3}\sw^4+{\Mb^4\over\MZ^4}
(24\sw^2-16\sw^4)-9{\Mb^2\over\MZ^2}
\nl&&\qquad\qquad{}
+{\alpha_s\over\pi}\Bigl(15-28\sw^2+{88\over 3}\sw^4\Bigr)\Bigr]
%\nl&&{}
%9{\Mb^2\over\MZ^2}\Bigl(2(-1+{4\over 3}\sw^2)^2-3.667\Bigr)+
%{\Mb^4\over\MZ^4}\Bigl(15-33(-1+{4\over 3}\sw^2)^2\Bigr)\Bigr)\Bigr]
\eeqar
and 
\beq
\GW ={\alpha\MW\over{2\sw^2}}\left [{3\over 2}+{\alpha_s\over\pi}
\right ] .
\eeq
For the strong coupling we use $\alpha_s=0.117$.

As to parton distributions, we have used CTEQ6M \cite{cteq} at the
following factorization scale:
\begin{equation}\label{eq:scaleWZ}
Q^2={1\over 2}\left (\MW^2+\MZ^2+\PT^2(l\bar\nu_l)+
\PT^2(l^\prime\bar{l^\prime})\right )
\end{equation}
This scale choice appears to be appropriate for the calculation of
differential cross sections, in particular for vector-boson
transverse-momentum distributions \cite{Dixon:1999di,Frixione:1992pj}.

We have, moreover, implemented a general set of cuts, proper for LHC 
analyses, defined as follows:
\begin{itemize}
\item {charged lepton transverse momentum $\PT(l)>20\GeV$},
  
\item {missing transverse momentum $\PTmiss> 20\GeV$},
  
\item {charged lepton pseudo-rapidity $|\eta_l |< 3$}, where
  $\eta_l=-\log\left (\tan(\theta_l/2)\right )$, and $\theta_l$ is the
  polar angle of particle $l$ with respect to the beam,
\item {lepton pair invariant mass $\PM (l^\prime\bar{l^\prime})\ge 0.201\GeV$.}
\end{itemize}
These cuts approximately simulate the detector acceptance.
For the processes considered, we have also implemented further 
cuts which are described in due time. In the following sections, we present 
results for the LHC at $\CM$ energy $\sqrt s=14\TeV$ and an integrated 
luminosity $L=100\fba^{-1}$.

\section{Gauge scheme and signal definition}
\label{sec:signal}

In this section we discuss how to identify and separate the signal of $\PW\PZ$ 
production from the background. Let us first define these two contributions 
diagrammatically.

\begin{figure}
  \unitlength 1cm
\begin{center}
\begin{picture}(16.,15.)
\put(-0.5,-4){\epsfig{file=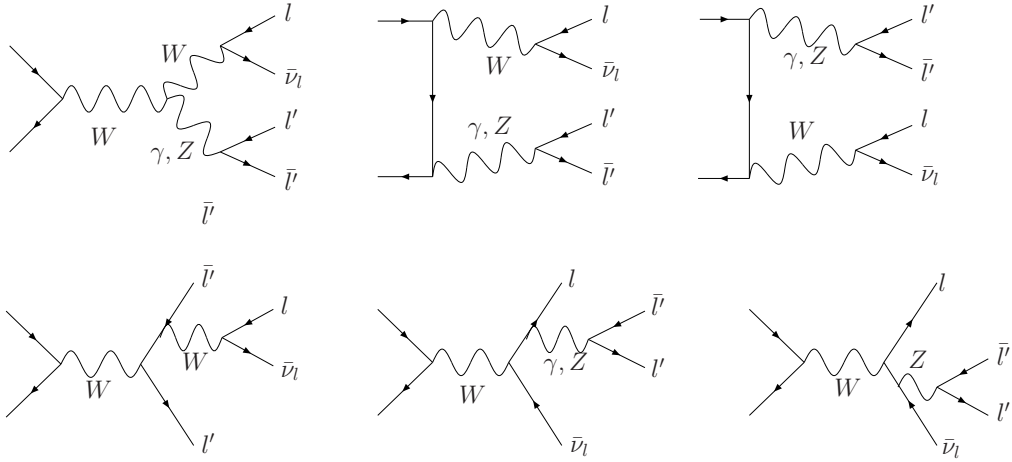,width=14cm}}
\end{picture}
\end{center}  
\vspace{-8.cm}
\caption{Feynman diagrams for the full process 
$\Pp\Pp\to l\bar\nu_ll^\prime\bar{l^\prime}$. The first row shows the 
doubly-resonant CC03 diagrams for $\PW\PZ$ production (when the photon is 
neglected). The latter represents the irreducible background, which the 
photon contribution in the first row must be added to.}
\label{fi:diagrams}
\end{figure}

The generic process $\Pp\Pp\to l\bar\nu_ll^\prime\bar{l^\prime}$ is described 
by the Feynman diagrams drawn in \reffi{fi:diagrams}. The three 
doubly-resonant graphs mediated by $\PW$ and $\PZ$-boson production are 
displayed in the first row. From LEP2 on, this is what we call CC03 signal for 
the di-boson production \footnote{
The CC03 cross section was introduced and discussed in \citeres{cc03_1,cc03_2} 
in order to extract the $\PW\PW$ signal from the full set of 
$e^+e^-\rightarrow 4f$ processes.}. The irreducible 
background, represented by singly-resonant and non-resonant diagrams, is 
instead shown in the second row, and partially in the first row by the graphs 
with virtual photon exchange.

From a practical point of view, the aim is to maximize the signal over the 
background ratio, picking out the kinematical regions where the first one is 
more enhanced and applying appropriate cuts to suppress the latter.
Hence, having at disposal a clear separation among the two contributions is 
highly desirable.

Unfortunately, signal and background do not individually preserve gauge 
invariance. Each of them includes indeed gauge-invariance-breaking terms which 
are only cancelled in their sum. Hence, only the full set of Feynman diagrams, 
i.e. the complete amplitude, is theoretically well-behaved.
Despite of that, one could still define a pseudo-observable using the pure
doubly-resonant CC03 contribution. 

The possibility of such a definition strongly depends on the size of the
terms which violate gauge invariance. As well known, their numerical impact 
can vary according to the energy and the off-shellness of the intermediate 
particles in the process. This generally makes their behaviour unpredictable. 
The off-shellness is indeed a variable one cannot always limit.
In the easy case at hand the virtuality of the produced $\PW$ and $\PZ$ 
bosons can be arbitrarily reduced, in this way suppressing the gauge violating 
terms. In the limit of on-shell $\PW\PZ$ production, the CC03 diagrams would 
in fact constitute a gauge independent set. But there are also opposite 
examples.
%this is not the most general case. 
When the virtual particles in a process are exchanged in t-channel, they 
cannot be forced anymore to be almost on-shell, leaving thus unconstrained the 
dangerous terms.

In the next two sections, we show that the gauge-fixing choice is the actual
control key for the gauge-invariance-breaking terms. 

\subsection{CC03 in Unitary Gauge}

The CC03 cross section was introduced at LEP2 in order to combine the different
final state measurements from the various collaborations, and increase the 
statistics. Usually defined either in the Unitary gauge or in the 
't Hooft-Feynman gauge, the CC03 cross section is not an observable, but at
LEP2 energies it was taken as a useful quantity. It contains interesting 
informations about triple gauge boson vertices, and is sensible to the 
$\PM_\PW$ value. At LEP2, the CC03 signal was then classified as a  
pseudo-observable and widely used. Its reliability was based on its closeness
to the full result, which implies neglegible gauge violating terms. 
But, the crucial caveat was that such a signal definition might become very 
problematic at future high energy colliders, owing to the much larger 
backgrounds and gauge dependences.  

In Unitary gauge, delicate cancellations between doubly-resonant (DR) and 
non-DR diagrams characterize the behaviour of off-shell cross sections in the 
high-energy regime. In the 't Hooft-Feynman gauge, this kind of cancellations 
generally appear moderately weakened, but still persist. In the massless limit 
we are working in, the two 
gauge schemes coincide. In the following, we refer to that as Unitary gauge.
    
\begin{figure}
  \begin{center}
\epsfig{file=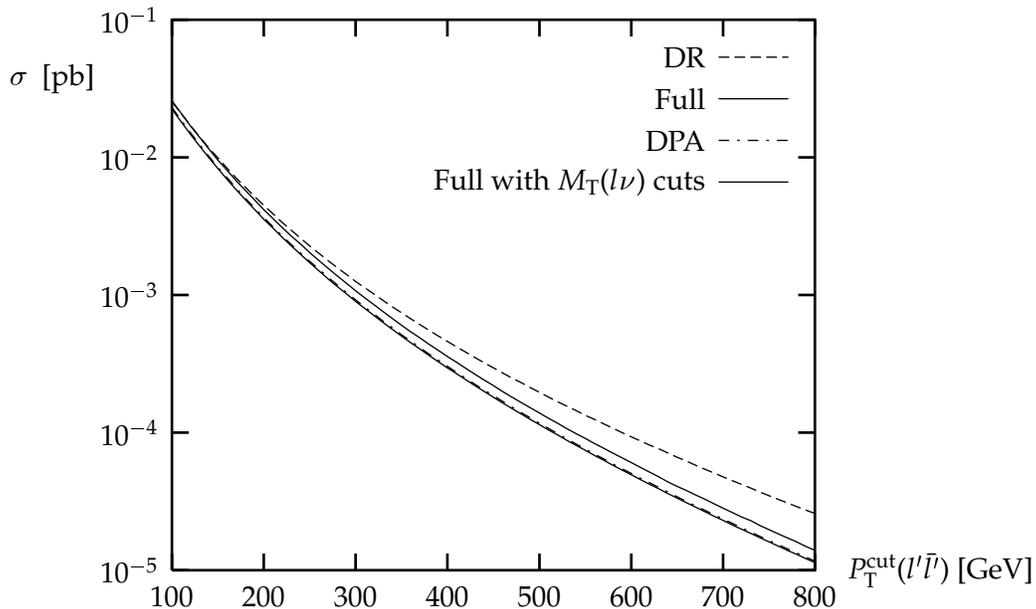, angle=0}
\end{center}  
\caption{Born cross section for the full process 
  $\Pp\Pp\to l\bar\nu_ll^\prime\bar{l^\prime}$ at $\sqrt{s}=14\TeV$ as a
  function of the cut on the transverse momentum of the reconstructed
  $\PZ$-boson. Standard cuts are applied. Legends as explained in the text.}
\label{fi:WZ_born_pt}
\end{figure}

For the example at hand, the behaviour of DR and non-DR diagrams is shown in 
\reffi{fi:WZ_born_pt} (see also \citere{adp}).
There, we have plotted the tree-level cross section as a function of the cut 
on the transverse momentum of the reconstructed $\PZ$-boson, $\PT(l'\bar l')$.
This cut selects large di-boson center-of-mass energies and wide scattering
angles of the produced vector-bosons. This is exactly the kinematical region 
dominated by the longitudinal gauge-boson production, and thus 
particularly sensitive to the gauge-violation effects we want to analyse.
 
The first two curves in \reffi{fi:WZ_born_pt} represent, from top to bottom, 
the contribution of the pure doubly-resonant diagrams, and the full result 
including all Feynman diagrams which contribute to the same final state.
The first clear information one can extract from the plot is that the 
DR contribution ($\Pp\Pp\to \PW\PZ\to 4f$), which is lower than the exact 
result ($\Pp\Pp\to 4f$) by about 1$\%$ around 
threshold, increases with energy relatively to the full result. For 
$\PTcut(l'\bar l')=300\GeV$, the difference between the two cross
sections is already of order 20$\%$, and at very large energies the DR
diagrams can even overestimate the result by a factor 2 or more.
%For $\PTcut(l'\bar l')=800\GeV$, DR and full cross sections are 
%$\sigma_{\DR}=2.4\times 10^{-5}\pba$ and 
%$\sigma_{\full}=1.4\times 10^{-5}\pba$ respectively.
   
This behavior can only be explained with the existence of strong interferences
between DR and non-DR subsets of diagrams, which are not separately 
gauge-independent. The consequence is that in Unitary gauge it is extremely 
hard to consider the pure DR contribution as a pseudo-observable.
The diagrammatic approach, commonly adopted at LEP2 for 
$\PW\PW$ and $\PZ\PZ$ physics, fails when describing the di-boson 
production at the LHC in the high-$\PT$ region.
In this gauge scheme, the only sensible observable is the total contribution.
%for it is the only well defined gauge-independent quantity. 
Thus, any signal definition seems to be completely lost.
A method which has been proposed to recover it consists in the double-pole 
approximation (DPA). This approximation emerges from the CC03 diagrams upon
projecting the vector-boson momenta in the matrix element to their on-shell 
values. This means that the DPA is based on the residue of the double 
resonance, which is related to the sub-processes of on-shell di-boson 
production and subsequent on-shell vector-boson decay. Owing to that, the DPA 
shares the property of preserving gauge invariance. 

For the case at hand, the DPA cross section is shown in \reffi{fi:WZ_born_pt},
where it is represented by the third curve from top to bottom. One can see
that the DPA is lower than the total cross section, their difference amounting 
%The difference between the exact result and the DPA amounts 
to roughly -15$\%$ for $\PT(l'\bar l')$ cuts above $100\GeV$. As displayed by
the solid line overlapping the DPA curve, the gap reduces to the per-cent 
level if one imposes cuts on the masses of the two lepton-pairs: 
$|\PM (l^\prime\bar{l^\prime})-\PM_\PZ |\le 20\GeV$ and  
$\MT(l\bar\nu_l)=\sqrt{\ET^2(l\bar\nu_l)-\PT^2(l\bar\nu_l)}\le \PM_\PW +20\GeV$.
This shows that, in contrast to the CC03 cross section in Unitary gauge, the 
DPA is theoretically well defined, and might be considered as a good estimate 
of the di-boson production when restricted to the doubly-resonant region.

However, the method has two substantial limitations. A first obvious price to 
be paid is the exclusion of the kinematical regions outside the s-channel 
resonances, where the DPA is not valid. The second limit is represented by the 
processes where the intermediate resonant particles are exchanged in t-channel 
(i.e. they have space-like virtuality). In this case, the DPA cannot be 
applied. One can rely only on its analogous given by the Equivalent Vector 
Boson approximation (EVBA), whose reliability is still debated for it crucially
depends on the applied kinematical cuts.

The two approximations can give an estimate of the signal. But their goodness
must be first checked against an exact computation. This bottom-top procedure 
of bringing the easier tool, represented by the approximate signal, to match 
the full result proves to be extremely powerful in some case. The DPA has 
been successfully employed for evaluating higher order corrections to the full 
process. But, also the reverse can be highly useful. That means defining an 
a-priori signal, which the full result should converge to. This is essential
in experiments. A scan of the full phase space 
in order to see where the signal is more enhanced, and how the background can
be suppressed might in fact be decisive for data analyses. 

In both Unitary and 't Hooft-Feynman gauge schemes this is not possible.
In the next section, we show that the Axial gauge is the appropriate
framework to obtain an independent and well-behaved signal definition.

\subsection{CC03 in Axial Gauge}
 
We consider the same process as before, 
$\Pp\Pp\to l\bar\nu_ll^\prime\bar{l^\prime}$, in Axial gauge. The matrix 
element is written according to the Feynman rules written in 
\refapp{sec:feynmanrules} (see also \citere{dk}). 

Since we assume all fermions to be massless, the contribution of the 
unphysical fields $\phi_\PZ$ and $\phi_\PW$ to the amplitude can be ignored.
This simplifies the computation sensibly, but does not alter the generality
of the results. The cross sections and distributions presented in this 
section have been obtained using $n_\mu =(2,1,1,1)$ as gauge vector.
However, we checked that the non-gauge-invariant quantities we analyse in the 
following have a very little dependence on that. 

Our aim is to show that in Axial gauge the diagrammatic approach can be 
recovered. This means that a signal, \ie a selected subset of diagrams, 
can be considered as a pseudo-observable even if non-gauge-invariant. 
To this end, we have chosen a phase-space region characterized by large
center of mass energies and large scattering angles of the produced vector 
bosons, $\PT(l^\prime\bar{l^\prime})> 800\GeV$. This is in fact the 
kinematical domain where the gauge-violating terms, if there, would be 
enhanced as displayed in \reffi{fi:WZ_born_pt} for the Unitary gauge.

\begin{figure}
  \unitlength 1cm
  \begin{center}
  \begin{picture}(16.,15.)
  \put(-2.5,-1){\epsfig{file=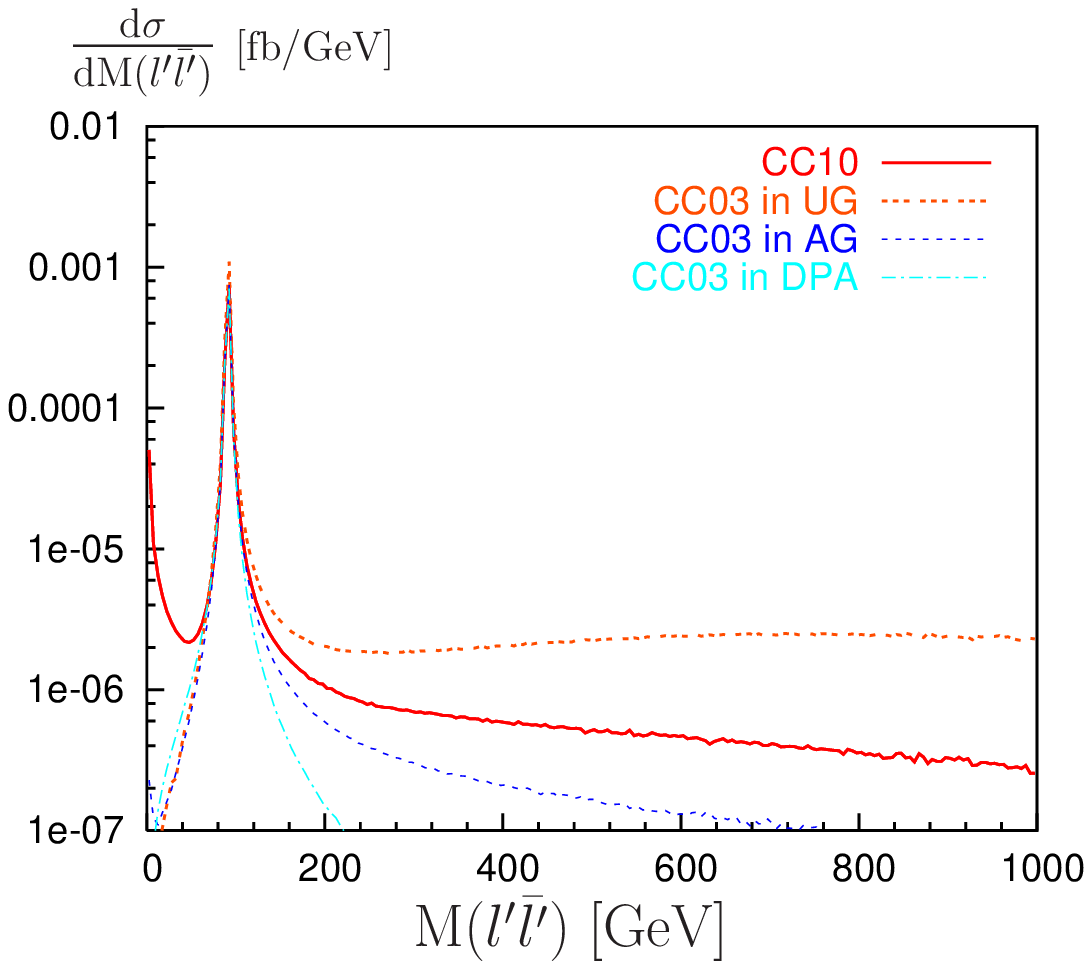,width=14cm}}
  \put(5.5,-1){\epsfig{file=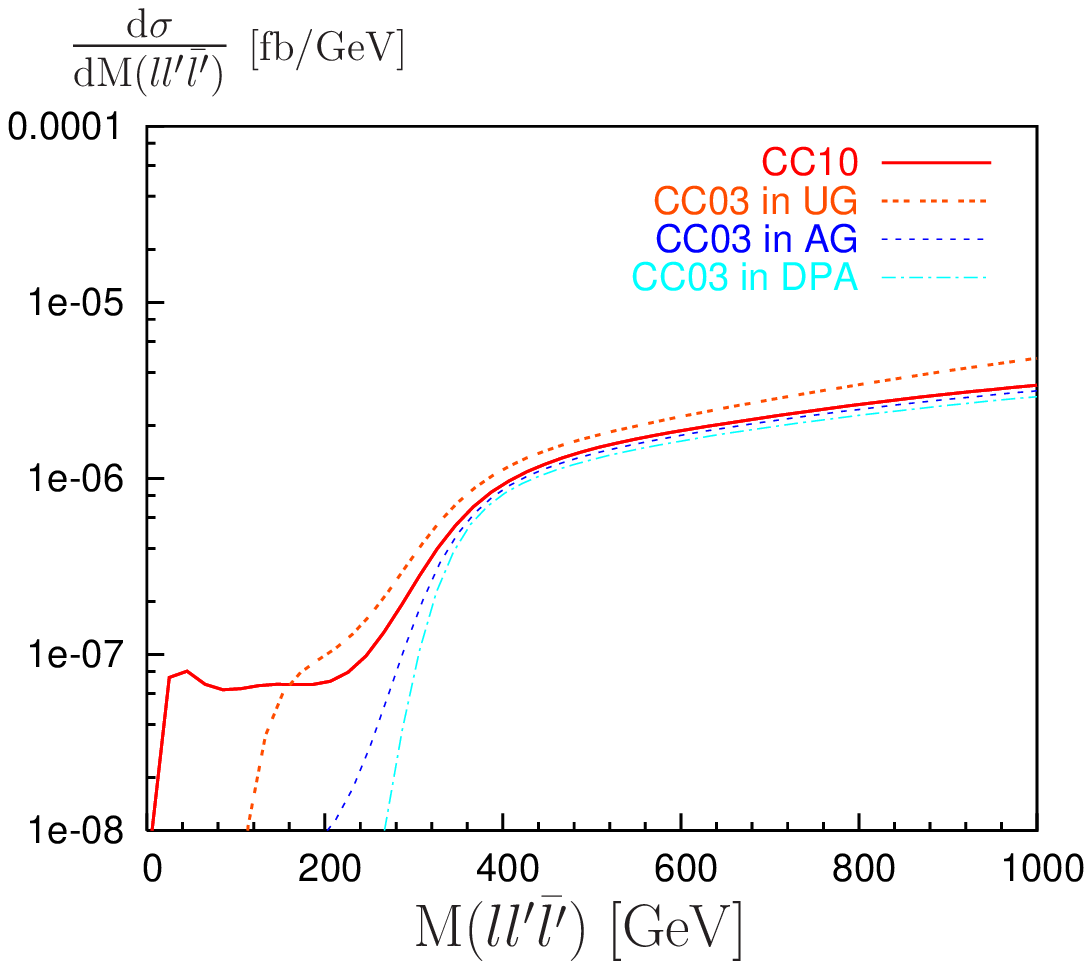,width=14cm}}
  \put(-2.5,-8){\epsfig{file=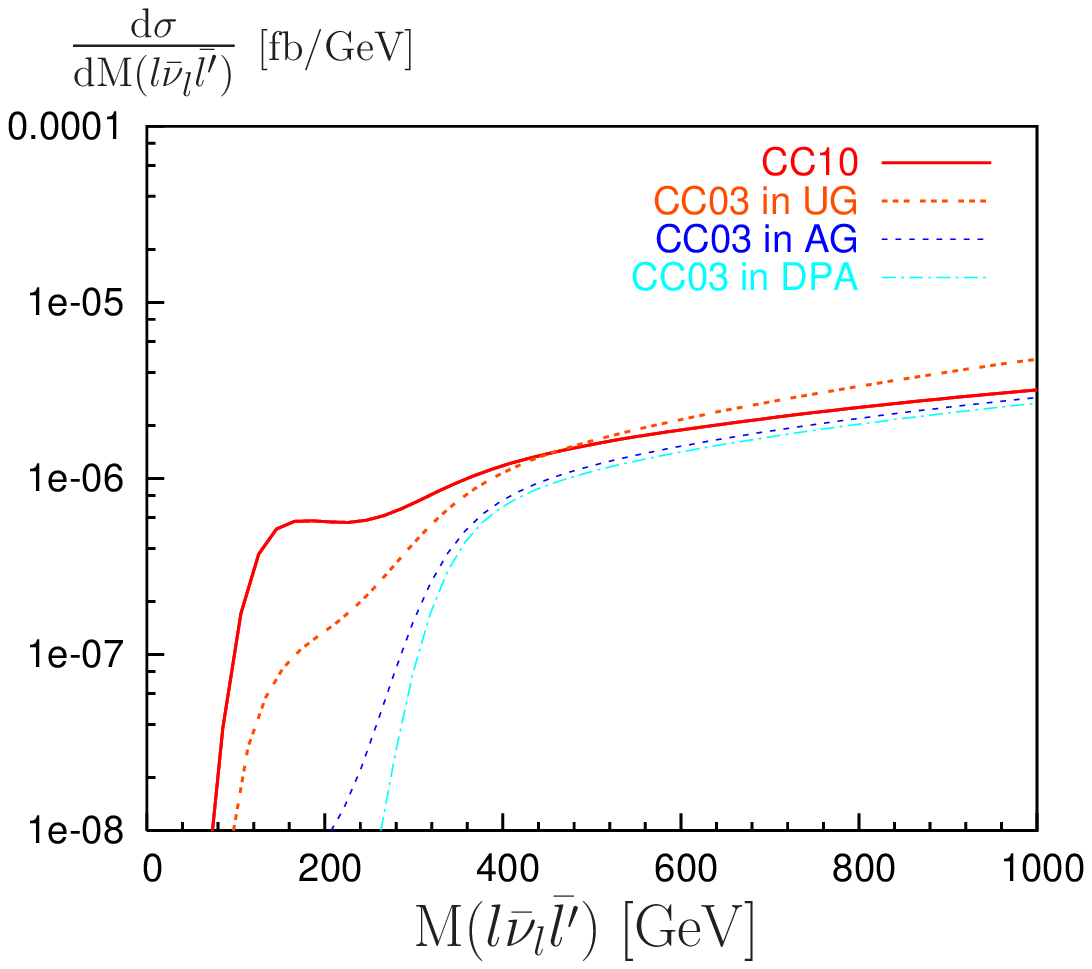,width=14cm}}
  \put(5.5,-8){\epsfig{file=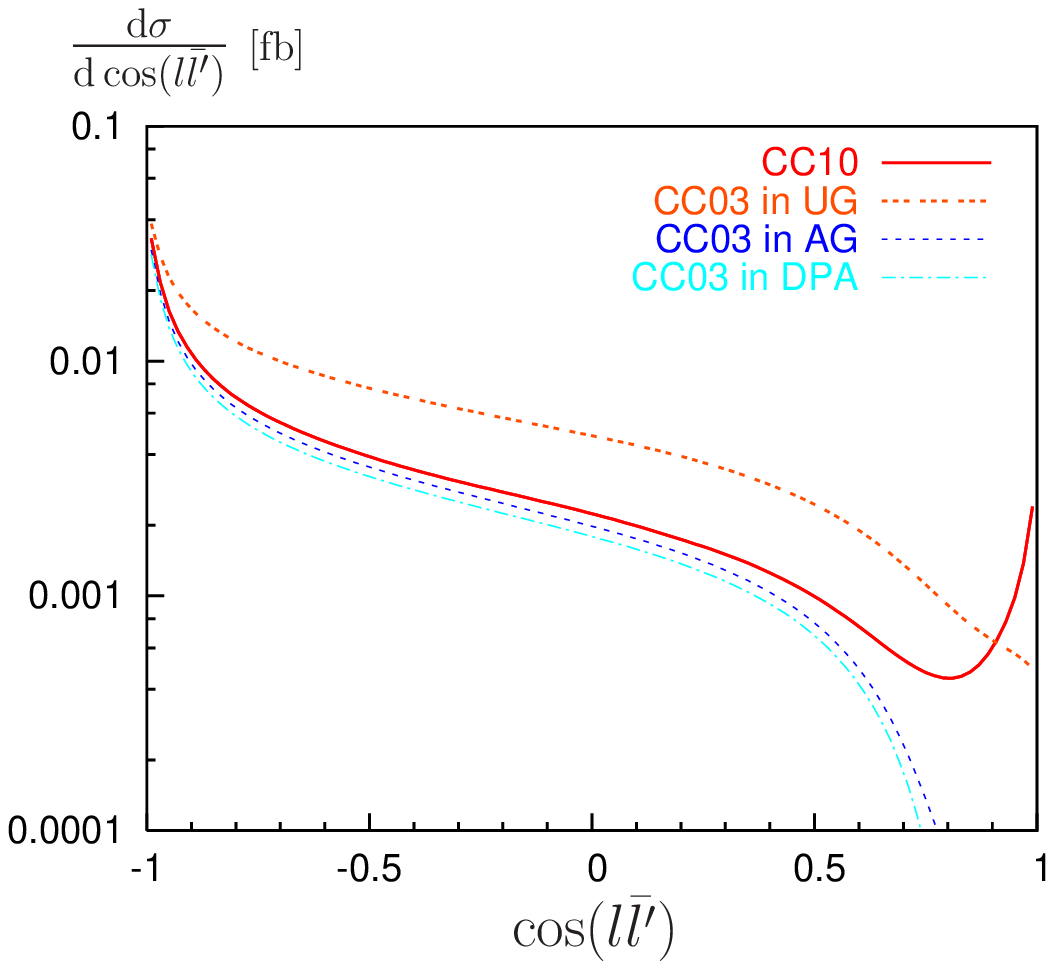,width=14cm}}
  \end{picture}
  \end{center}
\vspace{-2.cm}
\caption{Distributions for $\PW\PZ$ production. 
  (a) Invariant mass of the same-flavour charged lepton pair. (b) Invariant 
  mass of the charged leptons. (c) Invariant mass of the lepton pair which 
  might come from a $\PW$-boson, and the opposite-charge lepton which might 
  come from a $\PZ$-boson. (d) Cosine of the angle between opposite-charge 
  and different-flavour leptons.
  The contributions of the four final states 
  $l\bar\nu_ll^\prime\bar{l^\prime}$ where $l,l^\prime =e,\mu$
  are summed up, and standard cuts as well as 
  $\PT(\PZ)> 800\GeV$ are applied. Legends as explained in the text. 
}
\label{fi:WZ_s1}
\end{figure}

We have moreover selected four weakly correlated variables which reflect
our most direct expectations on the signal and background behaviour, namely:

\begin{itemize}
\item {$\PM(l^\prime\bar{l^\prime})$ - the invariant mass of the lepton pair 
which could come from the $\PZ$-boson decay,}
\item {$\PM(ll^\prime\bar{l^\prime})$ - the invariant mass of the three
charged leptons,}
\item{$\PM(l\bar\nu_l\bar{l^\prime})$ - the invariant mass of the two leptons
which could come from the $\PW$-boson decay plus the opposite-sign lepton
coming from the $\PZ$-boson,}
\item{$\cos (l\bar{l^\prime})$ - the cosine of the angle between the charged
lepton coming from the $\PW$-boson and the opposite-sign lepton from 
the $\PZ$-boson.}
\end{itemize}
The corresponding differential cross sections are plotted in \reffi{fi:WZ_s1}.
There, the solid line represents the full contribution coming from all ten
tree-level diagrams. The two dashed lines compare instead the CC03 signal
in the two gauge schemes. The long-dashed line gives CC03 in Unitary gauge (or 
t'Hooft-Feynman gauge), while the dashed one shows the CC03 signal in Axial 
gauge. Finally, as a reference, the dot-dashed line displays the DPA result.  
The first left-side plot contains the $\PM(l^\prime\bar{l^\prime})$ invariant 
mass. This observable peaks on $\PM_\PZ$, receiving the dominant contribution
from the CC03 signal. But, it is also expected to have some tail outside the 
resonant region, owing to all $\PW$-singly-resonant and non-resonant diagrams 
drawn in \reffi{fi:diagrams}. In particular, the photon exchange should 
generate a rise at low invariant mass values. The expected behaviour is well
reproduced by the solid line, as it must. 

Expectations are equally satisfied if one looks at the CC03 signal in Axial 
gauge. The so-defined signal is indeed concentrated around the 
$\PZ$-resonance, going sharply to zero beyond a few $\Gamma_\PZ$. The 
gauge-invariant DPA result comes as a further confirmation of the well-behaved 
CC03 in Axial gauge. The gauge-violating terms thus appear under control. 

Compared to these results, the Unitary gauge shows its ill-defined nature. In 
this scheme, the CC03 signal presents a long tail at large invariant masses, 
which is completely unphysical. It in fact stands up over the full result by 
an order of magnitude, implying the presence of huge gauge cancellations 
between the CC03 diagrams and the rest. The $\PZ$-boson invariant mass 
distribution clearly shows that the presence of the gauge-violating terms is 
strictly linked to the virtuality of the intermediate gauge boson.   

An analogous discussion holds for the other two invariant mass distributions
plotted in \reffi{fi:WZ_s1}. The top-right-side and the bottom-left-side plots
represent the momentum of the fermion propagator in graphs 5 and 4 of   
\reffi{fi:diagrams}, respectively. The two differential cross sections get the 
dominant contributions from the large masses, as the $\PW$ and $\PZ$ bosons 
are produced back-to-back mainly. But, they have also a sizeable low-mass 
component coming from the afore-mentioned singly-resonant diagrams, as shown
by the solid line. Once again, the CC03 signal in Axial gauge matches the
expectations, while the Unitary gauge gives a result which lies above the total
differential cross section, and displays unphysical tails.    

The last distribution on the bottom-right-side shows the cosine of the angle
between the two charged leptons which could come from the $\PW$ and the 
$\PZ$ bosons. The CC03 signal in Axial gauge peaks in the backward direction,
as the $\PW$ and $\PZ$ bosons are produced back-to-back. The full CC10 shows
in addition a forward rise coming from the graphs 4 and 5 in 
\reffi{fi:diagrams}. These contributions are in fact enhanced when the two 
leptons are produced collinearly. The Unitary gauge shows once more the usual
effect. It distorts and overestimates both the signal and the total 
differential cross sections, as for the previous variables.

\section{Conclusions}
\label{sec:conclusions}

In this letter, we have applied for the first time the Axial gauge \`a la Dams 
and Kleiss to analyze the $\SM$ physics at the LHC.
For a precise understanding of the high-energy phenomenology, having at hand
an unambiguous separation of signal and background is mandatory. We have shown 
that the Axial gauge is the appropriate framework to obtain a 
quasi-gauge-invariant signal definition. It allows in fact to isolate the 
signal trasparently, keeping the gauge-violating terms well under control even 
at very high energy scales.

For this first application, we have chosen to analyse the well-stated $\PW\PZ$ 
production process (a more complicated process mediated by vector boson 
scattering will be discussed in \citere{vbs_dk}). The signal definition has in 
this case a very well known
reference, namely the CC03 (NC02) cross section which was introduced and 
widely used at LEP2 for $\PW\PW$ ($\PZ\PZ$) physics. This quantity is not 
gauge invariant. It contains gauge-violating terms which cancel only when
summed up with the irreducible-background counterpart. Nevertheless,
it can be taken as a useful pseudo-observable if the gauge dependence is kept 
well below the experimental accuracy. 

That was the case at LEP2 energies, where the gauge-violating-breaking terms
were probed to be generally unimportant when computed in the 't Hooft-Feynman 
gauge. With increasing the energy, the size of the potentially badly-behaving 
terms might grow dramatically. Strong gauge cancellations between the various 
diagrams contributing to the same final state can take place, making any 
signal selection senseless.   

We have shown that the Axial gauge can recover the diagrammatic approach, and
give a well-behaved signal definition over the full energy domain.
   
\section*{Acknowledgements}
G.~Passarino is gratefully acknowledged for valuable comments on the 
subject treated and for carefully reading the manuscript. 
This work was supported by the Italian Ministero dell'Istruzione, 
dell'Universit\`a e della Ricerca (MIUR) under contract Decreto MIUR 
26-01-2001 N.13 ``Incentivazione alla mobilit\`a di studiosi stranieri ed 
italiani residenti all'estero''. 

\appendix

\section*{Appendix}

\section{Feynman rules}
\label{sec:feynmanrules}

In this appendix, we list the SM Feynman rules in Axial gauge. We adopt the 
same conventions as in \citere{dk}, which are here below summarized:
\begin{enumerate}
\item The Feynman rules that involve fermions are written only for the first 
      generation of leptons ($\Pe,\nu_\Pe$).
\item Particles and anti-particles are represented by lines with an arrow.
      The momentum flows in the direction of the arrow. For particles 
      described by lines without arrow, the momentum flows towards the vertex.
\item We use the following notation:
        \begin{center}
	$\Pg_\Pw = \frac{\Pg_\Pe}{\sin\theta_\Pw};\ \
	\Pg_\Pz = \frac{\Pg_\Pe}{\sin\theta_\Pw\cos\theta_\Pw};\ \
	p_l ={1\over 2}(1-\gamma^5);\ \
	p_r ={1\over 2}(1+\gamma^5).$
	\end{center}
\item If reversing all arrows on a vertex yields a different vertex,
	that vertex is also a vertex of the theory. 
	The corresponding vertex factor is obtained by conjugation of
        the original vertex, except for one factor of~$i$, and by reversing the
	sign of all momenta that belong to particles that do not carry an 
	arrow on their line. In the following, we give the expression of only 
	one sample vertex (the paired one must be derived).
\end{enumerate}

\subsection{Propagators}
\label{sec:propagators}
\fmruletab{
% 1
\begin{picture}(50,50)(0,0)
\Line(0,25)(50,25)\Text(25,26)[b]{$\scriptscriptstyle \PB(k)$}
\end{picture}
&
\frac{-i\left(
      g_{\nu\mu}
      -\frac{n_\nu k_\mu+n_\mu k_\nu}{n\cdot k}
      +k_\nu k_\mu\frac{n^2}{(n\cdot k)^2}
   \right)}{k^2+i\epsilon}
\cr
% 2
\begin{picture}(50,50)(0,0)
\ArrowLine(0,25)(50,25)\Text(25,26)[b]{$\scriptscriptstyle \PW(k)$}
\end{picture}
&
\frac{-i\left(
      g_{\nu\mu}
      -\frac{n_\nu k_\mu+n_\mu k_\nu}{n\cdot k}
      +k_\nu k_\mu\frac{n^2}{(n\cdot k)^2}
   \right)}{k^2-\PM_\PW^2+i\epsilon}
\cr
% 3
\begin{picture}(50,50)(0,0)
\ArrowLine(0,25)(50,25)\Text(25,27)[b]{$\scriptscriptstyle \phi_\PW(k)$}
\end{picture}
&
\frac i{k^2}
\cr
% 4
\begin{picture}(50,50)(0,0)
\Line(0,25)(50,25)\Text(25,26)[b]{$\scriptscriptstyle \PZ(k)$}
\end{picture}
&
\frac{-i\left(
      g_{\nu\mu}
      -\frac{n_\nu k_\mu+n_\mu k_\nu}{n\cdot k}
      +k_\nu k_\mu\frac{n^2}{(n\cdot k)^2}
   \right)}{k^2-\PM_\PZ^2+i\epsilon}
\cr
% 5
\begin{picture}(50,50)(0,0)
\Line(0,25)(50,25)\Text(25,26)[b]{$\scriptscriptstyle \phi_\PZ(k)$}
\end{picture}
&
\frac i{k^2}
\cr
% 6
\begin{picture}(50,50)(0,0)
\Line(0,25)(50,25)\Text(25,26)[b]{$\scriptscriptstyle \PH(k)$}
\end{picture}
&
\frac i{k^2-\PM_\PH^2+i\epsilon}
\cr
% 7
\begin{picture}(50,50)(0,0)
\ArrowLine(0,25)(50,25)\Text(25,27)[b]{$\scriptscriptstyle \Pe(k)$}
\end{picture}
&
\frac{i(\slash k+\Pm_\Pe)}{k^2-\Pm_\Pe^2+i\epsilon}
\cr
% 8
\begin{picture}(50,50)(0,0)
\ArrowLine(0,25)(50,25)\Text(25,27)[b]{$\scriptscriptstyle \nu_\Pe(k)$}
\end{picture}
&
\frac{i(\slash k+\Pm_{\nu_\Pe})}{k^2-\Pm_{\nu_\Pe}^2+i\epsilon}
\cr
% 9 zie 7
% 10 zie 8
}

\subsection{Triple boson couplings without Higgs}
\fmruletab{
% 11
\begin{picture}(50,50)(0,0)
\ArrowLine(0,0)(19,25)\Text(10,12)[lt]{$\scriptscriptstyle \PW(k_1)^\mu$}
\ArrowLine(19,25)(0,50)\Text(10,38)[lb]{$\scriptscriptstyle \PW(k_2)^\nu$}
\Line(19,25)(50,25)\Text(34,26)[b]{$\scriptscriptstyle \PB(k_3)^\sigma$}
\end{picture}
&
\eqalign{
i\Pg_\Pe\left[\vphantom{\frac12}\right.&
		g^{\nu\sigma}(k_2^\mu+k_3^\mu)
		+g^{\mu\sigma}(k_1^\nu-k_3^\nu)
		-g^{\mu\nu}(k_1^\sigma+k_2^\sigma)\cr
		&\left.-\PM_\PW^2\left(
			g^{\nu\sigma}\frac{k_1^\mu}{k_1^2}
			+g^{\mu\sigma}\frac{k_2^\nu}{k_2^2}
			-(k_1^\sigma+k_2^\sigma)\frac{k_1^\mu}{k_1^2}\frac{k_2^\nu}{k_2^2}
		\right)
	\right]
}
\cr
% 12
\begin{picture}(50,50)(0,0)
\ArrowLine(19,25)(0,0)\Text(10,12)[lt]{$\scriptscriptstyle \phi_\PW(k_1)$}
\ArrowLine(0,50)(19,25)\Text(10,38)[lb]{$\scriptscriptstyle \PW(k_2)^\mu$}
\Line(19,25)(50,25)\Text(34,26)[b]{$\scriptscriptstyle \PB(k_3)^\nu$}
\end{picture}
&
i\Pg_\Pe\PM_\PW\left(g^{\mu\nu}-\frac{(k_1^\nu+k_2^\nu)k_2^\mu}{k_2^2}\right)
\cr
% 13 zie 12
% 14
\begin{picture}(50,50)(0,0)
\ArrowLine(0,0)(19,25)\Text(10,12)[lt]{$\scriptscriptstyle \phi_\PW(k_1)$}
\ArrowLine(19,25)(0,50)\Text(10,38)[lb]{$\scriptscriptstyle \phi_\PW(k_2)$}
\Line(19,25)(50,25)\Text(34,26)[b]{$\scriptscriptstyle \PB(k_3)^\mu$}
\end{picture}
&
i\Pg_\Pe(k_1^\mu+k_2^\mu)
\cr
% 15
\begin{picture}(50,50)(0,0)
\ArrowLine(0,0)(19,25)\Text(10,12)[lt]{$\scriptscriptstyle \PW(k_2)^\nu$}
\ArrowLine(19,25)(0,50)\Text(10,38)[lb]{$\scriptscriptstyle \PW(k_3)^\sigma$}
\Line(19,25)(50,25)\Text(34,26)[b]{$\scriptscriptstyle \PZ(k_1)^\mu$}
\end{picture}
&	\eqalign{
	i\Pg_\Pw\cos\theta_\Pw\left[\vphantom{\frac12}\right.&
	-g^{\nu\sigma}(k_2^\mu+k_3^\mu)
	-g^{\mu\nu}(k_1^\sigma-k_2^\sigma)
	+g^{\mu\sigma}(k_1^\nu+k_3^\nu)\cr
	&+\PM_\PZ^2\sin^2\theta_\Pw\left(
		g^{\mu\sigma}\frac{k_2^\nu}{k_2^2}
		+g^{\mu\nu}\frac{k_3^\sigma}{k_3^2}
	\right)\cr
	&+\frac12\PM_\PZ^2\left(
		-(k_1^\sigma-k_2^\sigma)\frac{k_1^\mu}{k_1^2}\frac{k_2^\nu}{k_2^2}
		-(k_1^\nu+k_3^\nu)\frac{k_1^\mu}{k_1^2}\frac{k_3^\sigma}{k_3^2}
	\right)\cr
	&\left.+\PM_\PZ^2\left(\frac12-\sin^2\theta_\Pw\right)(k_2^\mu+k_3^\mu)
										\frac{k_2^\nu}{k_2^2}\frac{k_3^\sigma}{k_3^2}
	\right]\cr
	}
\cr
% 16
\begin{picture}(50,50)(0,0)
\ArrowLine(19,25)(0,0)\Text(10,12)[lt]{$\scriptscriptstyle \phi_\PW(k_2)$}
\ArrowLine(0,50)(19,25)\Text(10,38)[lb]{$\scriptscriptstyle \PW(k_3)^\nu$}
\Line(19,25)(50,25)\Text(34,26)[b]{$\scriptscriptstyle \PZ(k_1)^\mu$}
\end{picture}
&
-i\Pg_\Pz\PM_\PW\left(
	\sin^2\theta_\Pw g^{\mu\nu}
	-\frac12\frac{(k_1^\nu+k_2^\nu)k_1^\mu}{k_1^2}
	+\left(\cos^2\theta_\Pw-\frac12\right)\frac{(k_2^\mu+k_3^\mu)k_3^\nu}{k_3^2}
\right)
\cr
% 17 zie 16
% 18
\begin{picture}(50,50)(0,0)
\ArrowLine(0,0)(19,25)\Text(10,12)[lt]{$\scriptscriptstyle \phi_\PW(k_2)$}
\ArrowLine(19,25)(0,50)\Text(10,38)[lb]{$\scriptscriptstyle \phi_\PW(k_3)$}
\Line(19,25)(50,25)\Text(34,26)[b]{$\scriptscriptstyle \PZ(k_1)^\mu$}
\end{picture}
&
i\Pg_\Pw\left(k_2^\mu+k_3^\mu\right)\left(\cos\theta_\Pw-\frac1{2\cos\theta_\Pw}\right)
\cr
% 19
\begin{picture}(50,50)(0,0)
\ArrowLine(0,0)(19,25)\Text(10,12)[lt]{$\scriptscriptstyle \PW(k_2)^\mu$}
\ArrowLine(19,25)(0,50)\Text(10,38)[lb]{$\scriptscriptstyle \PW(k_3)^\nu$}
\Line(19,25)(50,25)\Text(34,26)[b]{$\scriptscriptstyle \phi_\PZ(k_1)$}
\end{picture}
&
\frac12\Pg_\Pw\PM_\PW\left(
	\frac{(k_2^\nu-k_1^\nu)k_2^\mu}{k_2^2}
	-\frac{(k_1^\mu+k_3^\mu)k_3^\nu}{k_3^2}
\right)
\cr
% 20
\begin{picture}(50,50)(0,0)
\ArrowLine(19,25)(0,0)\Text(10,12)[lt]{$\scriptscriptstyle \phi_\PW(k_2)$}
\ArrowLine(0,50)(19,25)\Text(10,38)[lb]{$\scriptscriptstyle \PW(k_3)^\mu$}
\Line(19,25)(50,25)\Text(34,26)[b]{$\scriptscriptstyle \phi_\PZ(k_1)$}
\end{picture}
&
\frac12\Pg_\Pw\left(k_1^\mu+k_2^\mu\right)
\cr
% 21 zie 20
}

\subsection{Triple boson couplings with Higgs}
\fmruletab{
% 22
\begin{picture}(50,50)(0,0)
\ArrowLine(0,0)(19,25)\Text(10,12)[lt]{$\scriptscriptstyle \PW(k_2)^\mu$}
\ArrowLine(19,25)(0,50)\Text(10,38)[lb]{$\scriptscriptstyle \PW(k_3)^\nu$}
\Line(19,25)(50,25)\Text(34,26)[b]{$\scriptscriptstyle \PH(k_1)$}
\end{picture}
&
\frac i2\Pg_\Pw\PM_\PW\left(
	2g^{\mu\nu}
	-\frac{k_3^\nu(k_1^\mu+k_3^\mu)}{k_3^2}
	+\frac{k_2^\mu(k_1^\nu-k_2^\nu)}{k_2^2}
	-\PM_\PH^2\frac{k_2^\mu}{k_2^2}\frac{k_3^\nu}{k_3^2}
\right)
\cr
% 23 zie 24
% 24
\begin{picture}(50,50)(0,0)
\ArrowLine(0,0)(19,25)\Text(10,12)[lt]{$\scriptscriptstyle \phi_\PW(k_2)$}
\ArrowLine(19,25)(0,50)\Text(10,38)[lb]{$\scriptscriptstyle \PW(k_3)^\mu$}
\Line(19,25)(50,25)\Text(34,26)[b]{$\scriptscriptstyle \PH(k_1)$}
\end{picture}
&
\frac i2\Pg_\Pw\left(k_2^\mu-k_1^\mu+\frac{\PM_\PH^2}{k_3^2}k_3^\mu\right)
\cr
% 25 
\begin{picture}(50,50)(0,0)
\ArrowLine(0,0)(19,25)\Text(10,12)[lt]{$\scriptscriptstyle \phi_\PW(k_2)$}
\ArrowLine(19,25)(0,50)\Text(10,38)[lb]{$\scriptscriptstyle \phi_\PW(k_3)$}
\Line(19,25)(50,25)\Text(34,26)[b]{$\scriptscriptstyle \PH(k_1)$}
\end{picture}
&
-\frac i2\Pg_\Pw\frac{\PM_\PH^2}{\PM_\PW}
\cr
% 26
\begin{picture}(50,50)(0,0)
\Line(0,0)(19,25)\Text(10,12)[lt]{$\scriptscriptstyle \PZ(k_2)^\mu$}
\Line(19,25)(0,50)\Text(10,38)[lb]{$\scriptscriptstyle \PZ(k_3)^\nu$}
\Line(19,25)(50,25)\Text(34,26)[b]{$\scriptscriptstyle \PH(k_1)$}
\end{picture}
&
i\Pg_\Pz\PM_\PZ\left(
	g^{\mu\nu}
	+\frac12(k_1^\nu-k_2^\nu)\frac{k_2^\mu}{k_2^2}
	+\frac12(k_1^\mu-k_3^\mu)\frac{k_3^\nu}{k_3^2}
	+\frac12\PM_\PH^2\frac{k_2^\mu}{k_2^2}\frac{k_3^\nu}{k_3^2}
\right)
\cr
% 27
\begin{picture}(50,50)(0,0)
\Line(0,0)(19,25)\Text(10,12)[lt]{$\scriptscriptstyle \phi_\PZ(k_2)$}
\Line(19,25)(0,50)\Text(10,38)[lb]{$\scriptscriptstyle \PZ(k_3)^\mu$}
\Line(19,25)(50,25)\Text(34,26)[b]{$\scriptscriptstyle \PH(k_1)$}
\end{picture}
&
\frac12\Pg_\Pz\left(k_1^\mu-k_2^\mu+\PM_\PH^2\frac{k_3^\mu}{k_3^2}\right)
\cr
% 28
\begin{picture}(50,50)(0,0)
\Line(0,0)(19,25)\Text(10,12)[lt]{$\scriptscriptstyle \phi_\PZ(k_2)$}
\Line(19,25)(0,50)\Text(10,38)[lb]{$\scriptscriptstyle \phi_\PZ(k_3)$}
\Line(19,25)(50,25)\Text(34,26)[b]{$\scriptscriptstyle \PH(k_1)$}
\end{picture}
&
-\frac i2\Pg_\Pz\frac{\PM_\PH^2}{\PM_\PZ}
\cr
% 29
\begin{picture}(50,50)(0,0)
\Line(0,0)(19,25)\Text(10,12)[lt]{$\scriptscriptstyle \PH(k_2)$}
\Line(19,25)(0,50)\Text(10,38)[lb]{$\scriptscriptstyle \PH(k_3)$}
\Line(19,25)(50,25)\Text(34,26)[b]{$\scriptscriptstyle \PH(k_1)$}
\end{picture}
&
-\frac{3i}2\Pg_\Pw\frac{\PM_\PH^2}{\PM_\PW}
\cr
}

\subsection{Couplings to the Fermions}
\fmruletab{
% 30
\begin{picture}(50,50)(0,0)
\ArrowLine(0,0)(19,25)\Text(10,12)[lt]{$\scriptscriptstyle \Pe(k_1)$}
\ArrowLine(19,25)(0,50)\Text(10,38)[lb]{$\scriptscriptstyle \Pe(k_2)$}
\Line(19,25)(50,25)\Text(34,26)[b]{$\scriptscriptstyle \PB(k_3)^\mu$}
\end{picture}
&
i\Pg_\Pe\gamma^\mu
\cr
% 31
\begin{picture}(50,50)(0,0)
\ArrowLine(0,0)(19,25)\Text(10,12)[lt]{$\scriptscriptstyle \Pe(k_1)$}
\ArrowLine(19,25)(0,50)\Text(10,38)[lb]{$\scriptscriptstyle \Pe(k_2)$}
\Line(19,25)(50,25)\Text(34,26)[b]{$\scriptscriptstyle \PZ(k_3)^\mu$}
\end{picture}
&
i\Pg_\Pz\left
(	\frac12\gamma^\mu p_l
	-\gamma^\mu\sin^2\theta_\Pw
	+\frac12\frac{\Pm_\Pe}{k_3^2}k_3^\mu\gamma^5\right
)
\cr
% 32
\begin{picture}(50,50)(0,0)
\ArrowLine(0,0)(19,25)\Text(10,12)[lt]{$\scriptscriptstyle \Pe(k_1)$}
\ArrowLine(19,25)(0,50)\Text(10,38)[lb]{$\scriptscriptstyle \Pe(k_2)$}
\Line(19,25)(50,25)\Text(34,26)[b]{$\scriptscriptstyle \phi_\PZ(k_3)$}
\end{picture}
&
\frac12\Pg_\Pz\frac{\Pm_\Pe}{\PM_\PZ}\gamma^5
\cr
% 33
\begin{picture}(50,50)(0,0)
\ArrowLine(0,0)(19,25)\Text(10,12)[lt]{$\scriptscriptstyle \Pe(k_1)$}
\ArrowLine(19,25)(0,50)\Text(10,38)[lb]{$\scriptscriptstyle \Pe(k_2)$}
\Line(19,25)(50,25)\Text(34,26)[b]{$\scriptscriptstyle \PH(k_3)$}
\end{picture}
&
-\frac i2\Pg_\Pw\frac{\Pm_\Pe}{\PM_\PW}
\cr
% 34
\begin{picture}(50,50)(0,0)
\ArrowLine(0,0)(19,25)\Text(10,12)[lt]{$\scriptscriptstyle \Pe(k_1)$}
\ArrowLine(19,25)(0,50)\Text(10,38)[lb]{$\scriptscriptstyle \nu_\Pe(k_2)$}
\ArrowLine(19,25)(50,25)\Text(34,26)[b]{$\scriptscriptstyle \PW(k_3)^\mu$}
\end{picture}
&
-\frac{i\Pg_\Pw}{\sqrt2}V_{11}\left(
	\gamma^\mu p_l
	+\left(\Pm_{\nu_\Pe}p_l-\Pm_\Pe p_r\right)\frac{k_3^\mu}{k_3^2}
\right)
\cr
% 35
\begin{picture}(50,50)(0,0)
\ArrowLine(0,0)(19,25)\Text(10,12)[lt]{$\scriptscriptstyle \Pe(k_1)$}
\ArrowLine(19,25)(0,50)\Text(10,38)[lb]{$\scriptscriptstyle \nu_\Pe(k_2)$}
\ArrowLine(19,25)(50,25)\Text(34,26)[b]{$\scriptscriptstyle \phi_\PW(k_3)$}
\end{picture}
&
\frac i{\sqrt2}\frac{\Pg_\Pw}{\PM_\PW}V_{11}\left(\Pm_{\nu_e}p_l-\Pm_\Pe p_r\right)
\cr
% 36 zie 34
% 37 zie 35
% 38 
\begin{picture}(50,50)(0,0)
\ArrowLine(0,0)(19,25)\Text(10,12)[lt]{$\scriptscriptstyle \nu_\Pe(k_1)$}
\ArrowLine(19,25)(0,50)\Text(10,38)[lb]{$\scriptscriptstyle \nu_\Pe(k_2)$}
\Line(19,25)(50,25)\Text(34,26)[b]{$\scriptscriptstyle \PZ(k_3)^\mu$}
\end{picture}
% 39 zie 30
&
-\frac i2\Pg_\Pz\left(\gamma^\mu p_l+\Pm_{\nu_\Pe}\gamma^5\frac{k_3^\mu}{k_3^2}\right)
\cr
% new rule
\begin{picture}(50,50)(0,0)
\ArrowLine(0,0)(19,25)\Text(10,12)[lt]{$\scriptscriptstyle \nu_\Pe(k_1)$}
\ArrowLine(19,25)(0,50)\Text(10,38)[lb]{$\scriptscriptstyle \nu_\Pe(k_2)$}
\Line(19,25)(50,25)\Text(34,26)[b]{$\scriptscriptstyle \phi_\PZ(k_3)$}
\end{picture}
&
-\frac12 \Pg_\Pw\frac{\Pm_{\nu_\Pe}}{\PM_\PW}\gamma^5
\cr
% new rule
\begin{picture}(50,50)(0,0)
\ArrowLine(0,0)(19,25)\Text(10,12)[lt]{$\scriptscriptstyle \nu_\Pe(k_1)$}
\ArrowLine(19,25)(0,50)\Text(10,38)[lb]{$\scriptscriptstyle \nu_\Pe(k_2)$}
\Line(19,25)(50,25)\Text(34,26)[b]{$\scriptscriptstyle \PH(k_3)$}
\end{picture}
&
-\frac i2\Pg_\Pw\frac{\Pm_{\nu_\Pe}}{\PM_\PW}
\cr
% new rule
\begin{picture}(50,50)(0,0)
\ArrowLine(19,25)(0,0)\Text(10,12)[lt]{$\scriptscriptstyle \mu(k_1)$}
\ArrowLine(0,50)(19,25)\Text(10,38)[lb]{$\scriptscriptstyle \nu_\Pe(k_2)$}
\ArrowLine(50,25)(19,25)\Text(34,27)[b]{$\scriptscriptstyle \PW(k_3)^\mu$}
\end{picture}
&
-\frac{i\Pg_\Pw}{\sqrt2}V^\dagger_{21}\left(
	\gamma^\mu p_l
	-\left(\Pm_\mu p_l-\Pm_{\nu_\Pe}p_r\right)\frac{k_3^\mu}{k_3^2}
\right)
\cr
% new rule
\begin{picture}(50,50)(0,0)
\ArrowLine(19,25)(0,0)\Text(10,12)[lt]{$\scriptscriptstyle \mu(k_1)$}
\ArrowLine(0,50)(19,25)\Text(10,38)[lb]{$\scriptscriptstyle \nu_\Pe(k_2)$}
\ArrowLine(50,25)(19,25)\Text(34,27)[b]{$\scriptscriptstyle \phi_\PW(k_3)$}
\end{picture}
&
-\frac{i}{\sqrt2}\frac{\Pg_\Pw}{\PM_\PW}V^\dagger_{21}\left(
	\Pm_\mu p_l
	-\Pm_{\nu_\Pe}p_r
\right)
\cr
% 40 zie 31
% 41 zie 32
% 42 zie 33
% 43 zie 34
% 44 zie 35
% 45 zie 34
% 46 zie 35
% 47 zie 38
}

\subsection{Quadruple boson couplings among $\PB$, $\PW$ and $\phi_\PW$}
\fmruletab{
% 48
\begin{picture}(50,50)(0,0)
\ArrowLine(0,0)(25,25)\Text(12,13)[rb]{$\scriptscriptstyle \PW(k_1)^\mu$}
\ArrowLine(25,25)(0,50)\Text(12,37)[rt]{$\scriptscriptstyle \PW(k_2)^\nu$}
\Line(50,0)(25,25)\Text(38,13)[lb]{$\scriptscriptstyle \PB(k_3)^\sigma$}
\Line(50,50)(25,25)\Text(38,37)[lt]{$\scriptscriptstyle \PB(k_4)^\tau$}
\end{picture}
&
i\Pg_\Pe^2\left(
	-2g^{\mu\nu}g^{\sigma\tau}
	+g^{\mu\sigma}g^{\nu\tau}
	+g^{\mu\tau}g^{\nu\sigma}
	+2\PM_\PW^2g^{\sigma\tau}\frac{k_1^\mu}{k_1^2}\frac{k_2^\nu}{k_2^2}\right)
\cr
% 49
\begin{picture}(50,50)(0,0)
\ArrowLine(0,0)(25,25)\Text(12,13)[rb]{$\scriptscriptstyle \PW(k_1)^\mu$}
\ArrowLine(0,50)(25,25)\Text(12,37)[rt]{$\scriptscriptstyle \PW(k_2)^\nu$}
\ArrowLine(25,25)(50,0)\Text(38,13)[lb]{$\scriptscriptstyle \PW(k_3)^\sigma$}
\ArrowLine(25,25)(50,50)\Text(38,37)[lt]{$\scriptscriptstyle \PW(k_4)^\tau$}
\end{picture}
&
\eqalign{
i\Pg_\Pw^2\left[\vphantom{\frac12}\right.&
	2g^{\mu\nu}g^{\sigma\tau}
	-g^{\mu\sigma}g^{\nu\tau}
	-g^{\mu\tau}g^{\nu\sigma}\cr
	&+\frac12\PM_\PW^2\left(
		g^{\nu\tau}\frac{k_1^\mu}{k_1^2}\frac{k_3^\sigma}{k_3^2}
		+g^{\nu\sigma}\frac{k_1^\mu}{k_1^2}\frac{k_4^\tau}{k_4^2}
		+g^{\mu\tau}\frac{k_2^\nu}{k_2^2}\frac{k_3^\sigma}{k_3^2}
		+g^{\mu\sigma}\frac{k_2^\nu}{k_2^2}\frac{k_4^\tau}{k_4^2}
	\right)\cr
	&\left.-\frac12\PM_\PW^2\PM_\PH^2\frac{k_1^\mu}{k_1^2}\frac{k_2^\nu}{k_2^2}
										\frac{k_3^\sigma}{k_3^2}\frac{k_4^\tau}{k_4^2}
\right]\cr
}
\cr
% 50
\begin{picture}(50,50)(0,0)
\ArrowLine(25,25)(0,0)\Text(12,13)[rb]{$\scriptscriptstyle \phi_\PW(k_1)$}
\ArrowLine(0,50)(25,25)\Text(12,37)[rt]{$\scriptscriptstyle \PW(k_2)^\mu$}
\Line(50,0)(25,25)\Text(38,13)[lb]{$\scriptscriptstyle \PB(k_3)^\nu$}
\Line(50,50)(25,25)\Text(38,37)[lt]{$\scriptscriptstyle \PB(k_4)^\sigma$}
\end{picture}
&
-2i\Pg_\Pe^2\PM_\PW g^{\nu\sigma}\frac{k_2^\mu}{k_2^2}
\cr
% 51
\begin{picture}(50,50)(0,0)
\ArrowLine(25,25)(0,0)\Text(12,13)[rb]{$\scriptscriptstyle \phi_\PW(k_1)$}
\ArrowLine(0,50)(25,25)\Text(12,37)[rt]{$\scriptscriptstyle \PW(k_2)^\mu$}
\ArrowLine(50,0)(25,25)\Text(38,13)[lb]{$\scriptscriptstyle \PW(k_3)^\nu$}
\ArrowLine(25,25)(50,50)\Text(38,37)[lt]{$\scriptscriptstyle \PW(k_4)^\sigma$}
\end{picture}
&
\frac i2\Pg_\Pw^2\PM_\PW\left(
	-g^{\nu\sigma}\frac{k_2^\mu}{k_2^2}
	-g^{\mu\sigma}\frac{k_3^\nu}{k_3^2}
	+\PM_\PH^2\frac{k_2^\mu}{k_2^2}\frac{k_3^\nu}{k_3^2}
		\frac{k_4^\sigma}{k_4^2}\right)
\cr
% 52
\begin{picture}(50,50)(0,0)
\ArrowLine(25,25)(0,0)\Text(12,13)[rb]{$\scriptscriptstyle \phi_\PW(k_1)$}
\ArrowLine(25,25)(0,50)\Text(12,37)[rt]{$\scriptscriptstyle \phi_\PW(k_2)$}
\ArrowLine(50,0)(25,25)\Text(38,13)[lb]{$\scriptscriptstyle \PW(k_3)^\mu$}
\ArrowLine(50,50)(25,25)\Text(38,37)[lt]{$\scriptscriptstyle \PW(k_4)^\nu$}
\end{picture}
&-\frac i2 \Pg_\Pw^2\PM_\PH^2\frac{k_3^\mu}{k_3^2}\frac{k_4^\nu}{k_4^2}
\cr
% 53 zie 50
% 54 zie 51
% 55
\begin{picture}(50,50)(0,0)
\ArrowLine(0,0)(25,25)\Text(12,13)[rb]{$\scriptscriptstyle \phi_\PW(k_1)$}
\ArrowLine(25,25)(0,50)\Text(12,37)[rt]{$\scriptscriptstyle \phi_\PW(k_2)$}
\Line(50,0)(25,25)\Text(38,13)[lb]{$\scriptscriptstyle \PB(k_3)^\mu$}
\Line(50,50)(25,25)\Text(38,37)[lt]{$\scriptscriptstyle \PB(k_4)^\nu$}
\end{picture}
&
2i\Pg_\Pe^2g^{\mu\nu}
\cr
% 56
\begin{picture}(50,50)(0,0)
\ArrowLine(0,0)(25,25)\Text(12,13)[rb]{$\scriptscriptstyle \phi_\PW(k_1)$}
\ArrowLine(25,25)(0,50)\Text(12,37)[rt]{$\scriptscriptstyle \phi_\PW(k_2)$}
\ArrowLine(50,0)(25,25)\Text(38,13)[lb]{$\scriptscriptstyle \PW(k_3)^\mu$}
\ArrowLine(25,25)(50,50)\Text(38,37)[lt]{$\scriptscriptstyle \PW(k_4)^\nu$}
\end{picture}
&
\frac i2\Pg_\Pw^2\left(
	g^{\mu\nu}-\PM_\PH^2\frac{k_3^\mu}{k_3^2}\frac{k_4^\nu}{k_4^2}\right)
\cr
% 57
\begin{picture}(50,50)(0,0)
\ArrowLine(0,0)(25,25)\Text(12,13)[rb]{$\scriptscriptstyle \phi_\PW(k_1)$}
\ArrowLine(25,25)(0,50)\Text(12,37)[rt]{$\scriptscriptstyle \phi_\PW(k_2)$}
\ArrowLine(25,25)(50,0)\Text(38,13)[lb]{$\scriptscriptstyle \phi_\PW(k_3)$}
\ArrowLine(50,50)(25,25)\Text(38,37)[lt]{$\scriptscriptstyle \PW(k_4)^\mu$}
\end{picture}
&
\frac i2\Pg_\Pw^2\frac{\PM_\PH^2}{\PM_\PW}\frac{k_4^\mu}{k_4^2}
\cr
% 58 zie 52
% 59 zie 57
% 60
\begin{picture}(50,50)(0,0)
\ArrowLine(0,0)(25,25)\Text(12,13)[rb]{$\scriptscriptstyle \phi_\PW(k_1)$}
\ArrowLine(0,50)(25,25)\Text(12,37)[rt]{$\scriptscriptstyle \phi_\PW(k_2)$}
\ArrowLine(25,25)(50,0)\Text(38,13)[lb]{$\scriptscriptstyle \phi_\PW(k_3)$}
\ArrowLine(25,25)(50,50)\Text(38,37)[lt]{$\scriptscriptstyle \phi_\PW(k_4)$}
\end{picture}
&
-\frac i2\Pg_\Pw^2\frac{\PM_\PH^2}{\PM_\PW^2}
\cr
}

\subsection{Quadruple boson couplings with $\PZ$, and without $\phi_\PZ$ or 
$\PH$}
\fmruletab{
% 61
\begin{picture}(50,50)(0,0)
\Line(25,25)(0,0)\Text(12,13)[rb]{$\scriptscriptstyle \PZ(k_1)^\mu$}
\ArrowLine(0,50)(25,25)\Text(12,37)[rt]{$\scriptscriptstyle \PW(k_2)^\nu$}
\ArrowLine(25,25)(50,0)\Text(38,13)[lb]{$\scriptscriptstyle \PW(k_3)^\sigma$}
\Line(25,25)(50,50)\Text(38,37)[lt]{$\scriptscriptstyle \PB(k_4)^\tau$}
\end{picture}
&
\eqalign{
i\Pg_\Pe\Pg_\Pw\cos\theta_\Pw\left[
	-2g^{\mu\tau}g^{\nu\sigma}
	+g^{\mu\nu}g^{\sigma\tau}
	+g^{\mu\sigma}g^{\nu\tau}+\PM_\PZ^2\left(
		\frac12g^{\sigma\tau}\frac{k_1^\mu}{k_1^2}\frac{k_2^\nu}{k_2^2}\right.\right.\cr
	\left.\left.-\frac12g^{\nu\tau}\frac{k_1^\mu}{k_1^2}\frac{k_3^\sigma}{k_3^2}
		+g^{\mu\tau}(2\cos\theta_w^2-1)\frac{k_2^\nu}{k_2^2}
																	\frac{k_3^\sigma}{k_3^2}
	\right)
\right]\cr
}
\cr
% 62
\begin{picture}(50,50)(0,0)
\Line(25,25)(0,0)\Text(12,13)[rb]{$\scriptscriptstyle \PZ(k_1)^\mu$}
\ArrowLine(25,25)(0,50)\Text(12,37)[rt]{$\scriptscriptstyle \phi_\PW(k_2)$}
\ArrowLine(50,0)(25,25)\Text(38,13)[lb]{$\scriptscriptstyle \PW(k_3)^\nu$}
\Line(25,25)(50,50)\Text(38,37)[lt]{$\scriptscriptstyle \PB(k_4)^\sigma$}
\end{picture}
&
i\Pg_\Pe\Pg_\Pz\PM_\PW\left(
	\frac12g^{\nu\sigma}\frac{k_1^\mu}{k_1^2}
	+g^{\mu\sigma}(1-2\cos^2\theta_\Pw)\frac{k_3^\nu}{k_3^2}
\right)
\cr
% 63 zie 62
% 64
\begin{picture}(50,50)(0,0)
\Line(25,25)(0,0)\Text(12,13)[rb]{$\scriptscriptstyle \PZ(k_1)^\mu$}
\ArrowLine(0,50)(25,25)\Text(12,37)[rt]{$\scriptscriptstyle \phi_\PW(k_2)$}
\ArrowLine(25,25)(50,0)\Text(38,13)[lb]{$\scriptscriptstyle \phi_\PW(k_3)$}
\Line(25,25)(50,50)\Text(38,37)[lt]{$\scriptscriptstyle \PB(k_4)^\nu$}
\end{picture}
&
i\Pg_\Pe\Pg_\Pz(2\cos^2\theta_\Pw-1)g^{\mu\nu}
\cr
% 65
\begin{picture}(50,50)(0,0)
\Line(25,25)(0,0)\Text(12,13)[rb]{$\scriptscriptstyle \PZ(k_1)^\mu$}
\Line(0,50)(25,25)\Text(12,37)[rt]{$\scriptscriptstyle \PZ(k_2)^\nu$}
\ArrowLine(50,0)(25,25)\Text(38,13)[lb]{$\scriptscriptstyle \PW(k_3)^\sigma$}
\ArrowLine(25,25)(50,50)\Text(38,37)[lt]{$\scriptscriptstyle \PW(k_4)^\tau$}
\end{picture}
&
\eqalign{
-i\Pg_\Pw^2\left[\vphantom{\frac12}\right.
	&\cos^2\theta_\Pw\left(
		2g^{\mu\nu}g^{\sigma\tau}
		-g^{\mu\sigma}g^{\nu\tau}
		-g^{\mu\tau}g^{\nu\sigma}
	\right)\cr
	&+\frac12\PM_\PZ^2\sin^2\theta_\Pw\left(
		\frac1{\sin^2\theta_\Pw}g^{\sigma\tau}\frac{k_1^\mu}{k_1^2}
																		\frac{k_2^\nu}{k_2^2}
		+g^{\nu\tau}\frac{k_1^\mu}{k_1^2}\frac{k_3^\sigma}{k_3^2}
		-g^{\nu\sigma}\frac{k_1^\mu}{k_1^2}\frac{k_4^\tau}{k_4^2}\right.\cr
		&+g^{\mu\tau}\frac{k_2^\nu}{k_2^2}\frac{k_3^\sigma}{k_3^2}
		-g^{\mu\sigma}\frac{k_2^\nu}{k_2^2}\frac{k_4^\tau}{k_4^2}
		+\left(4\cos^2\theta_\Pw-\frac1{\sin^2\theta_\Pw}\right)g^{\mu\nu}
										\frac{k_3^\sigma}{k_3^2}\frac{k_4^\tau}{k_4^2}\cr
		&\left.\left.-\frac{\PM_\PH^2}{2\sin^2\theta_\Pw}\frac{k_1^\mu}{k_1^2}
				\frac{k_2^\nu}{k_2^2}\frac{k_3^\sigma}{k_3^2}\frac{k_4^\tau}{k_4^2}
	\right)
\right]\cr
}
\cr
% 66
\begin{picture}(50,50)(0,0)
\Line(25,25)(0,0)\Text(12,13)[rb]{$\scriptscriptstyle \PZ(k_1)^\mu$}
\Line(0,50)(25,25)\Text(12,37)[rt]{$\scriptscriptstyle \PZ(k_2)^\nu$}
\ArrowLine(25,25)(50,0)\Text(38,13)[lb]{$\scriptscriptstyle \phi_\PW(k_3)$}
\ArrowLine(50,50)(25,25)\Text(38,37)[lt]{$\scriptscriptstyle \PW(k_4)^\sigma$}
\end{picture}
&
\eqalign{
-\frac i2\frac{\Pg_\Pe^2\PM_\PW}{\cos^2\theta_\Pw}\left[\vphantom{\frac12}\right.
	g^{\nu\sigma}\frac{k_1^\mu}{k_1^2}
	+g^{\mu\sigma}\frac{k_2^\nu}{k_2^2}
	+\left(\frac1{\sin^2\theta_\Pw}-4\cos^2\theta_\Pw\right)g^{\mu\nu}
																	\frac{k_4^\sigma}{k_4^2}\cr
	\left.+\frac1{2\sin^2\theta_\Pw}\PM_\PH^2\frac{k_1^\mu}{k_1^2}
											\frac{k_2^\nu}{k_2^2}\frac{k_4^\sigma}{k_4^2}
\right]\cr
}
\cr
% 67 zie 66
% 68
\begin{picture}(50,50)(0,0)
\Line(25,25)(0,0)\Text(12,13)[rb]{$\scriptscriptstyle \PZ(k_1)^\mu$}
\Line(25,25)(0,50)\Text(12,37)[rt]{$\scriptscriptstyle \PZ(k_2)^\nu$}
\ArrowLine(50,0)(25,25)\Text(38,13)[lb]{$\scriptscriptstyle \phi_\PW(k_3)$}
\ArrowLine(25,25)(50,50)\Text(38,37)[lt]{$\scriptscriptstyle \phi_\PW(k_4)$}
\end{picture}
&
i\Pg_\Pz^2\left(
	\left(\frac12-2\cos^2\theta_\Pw\sin^2\theta_\Pw\right)g^{\mu\nu}
	+\frac14\PM_\PH^2\frac{k_1^\mu}{k_1^2}\frac{k_2^\nu}{k_2^2}
\right)
\cr
% 69
\begin{picture}(50,50)(0,0)
\Line(25,25)(0,0)\Text(12,13)[rb]{$\scriptscriptstyle \PZ(k_1)^\mu$}
\Line(25,25)(0,50)\Text(12,37)[rt]{$\scriptscriptstyle \PZ(k_2)^\nu$}
\Line(50,0)(25,25)\Text(38,13)[lb]{$\scriptscriptstyle \PZ(k_3)^\sigma$}
\Line(25,25)(50,50)\Text(38,37)[lt]{$\scriptscriptstyle \PZ(k_4)^\tau$}
\end{picture}
&
\eqalign{
	-\frac i2\Pg_\Pz^2\PM_\PZ^2\left(\vphantom{\frac12}\right.&
		g^{\mu\nu}\frac{k_3^\sigma}{k_3^2}\frac{k_4^\tau}{k_4^2}
		+g^{\mu\sigma}\frac{k_2^\nu}{k_2^2}\frac{k_4^\tau}{k_4^2}
		+g^{\mu\tau}\frac{k_2^\nu}{k_2^2}\frac{k_3^\sigma}{k_3^2}
		+g^{\nu\sigma}\frac{k_1^\mu}{k_1^2}\frac{k_4^\tau}{k_4^2}\cr
		&\left.+g^{\nu\tau}\frac{k_1^\mu}{k_1^2}\frac{k_3^\sigma}{k_3^2}
		+g^{\sigma\tau}\frac{k_1^\mu}{k_1^2}\frac{k_2^\nu}{k_2^2}
		+\frac32\PM_\PH^2\frac{k_1^\mu}{k_1^2}\frac{k_2^\nu}{k_2^2}
											\frac{k_3^\sigma}{k_3^2}\frac{k_4^\tau}{k_4^2}
	\right)
}
\cr
}

\subsection{Quadruple boson couplings with one $\phi_\PZ$ and no $\PH$}
\fmruletab{
% 70
\begin{picture}(50,50)(0,0)
\Line(25,25)(0,0)\Text(12,13)[rb]{$\scriptscriptstyle \phi_\PZ(k_1)$}
\ArrowLine(0,50)(25,25)\Text(12,37)[rt]{$\scriptscriptstyle \PW(k_2)^\mu$}
\ArrowLine(25,25)(50,0)\Text(38,13)[lb]{$\scriptscriptstyle \PW(k_3)^\nu$}
\Line(25,25)(50,50)\Text(38,37)[lt]{$\scriptscriptstyle \PB(k_4)^\sigma$}
\end{picture}
&
\frac12\Pg_\Pe\Pg_\Pw\PM_\PW\left(
	g^{\nu\sigma}\frac{k_2^\mu}{k_2^2}
	-g^{\mu\sigma}\frac{k_3^\nu}{k_3^2}
\right)
\cr
% 71
\begin{picture}(50,50)(0,0)
\Line(25,25)(0,0)\Text(12,13)[rb]{$\scriptscriptstyle \phi_\PZ(k_1)$}
\ArrowLine(25,25)(0,50)\Text(12,37)[rt]{$\scriptscriptstyle \phi_\PW(k_2)$}
\ArrowLine(50,0)(25,25)\Text(38,13)[lb]{$\scriptscriptstyle \PW(k_3)^\mu$}
\Line(25,25)(50,50)\Text(38,37)[lt]{$\scriptscriptstyle \PB(k_4)^\nu$}
\end{picture}
&
\frac12\Pg_\Pe\Pg_\Pw g^{\mu\nu}
\cr
% 72 zie 71
% 73
\begin{picture}(50,50)(0,0)
\Line(25,25)(0,0)\Text(12,13)[rb]{$\scriptscriptstyle \phi_\PZ(k_1)$}
\Line(25,25)(0,50)\Text(12,37)[rt]{$\scriptscriptstyle \PZ(k_2)^\mu$}
\ArrowLine(50,0)(25,25)\Text(38,13)[lb]{$\scriptscriptstyle \PW(k_3)^\nu$}
\ArrowLine(25,25)(50,50)\Text(38,37)[lt]{$\scriptscriptstyle \PW(k_4)^\sigma$}
\end{picture}
&
\frac12\Pg_\Pe^2\PM_\PZ\left(
	-\frac{g^{\nu\sigma}}{\sin^2\theta_w}\frac{k_2^\mu}{k_2^2}
	-g^{\mu\sigma}\frac{k_3^\nu}{k_3^2}
	+g^{\mu\nu}\frac{k_4^\sigma}{k_4^2}
	+\frac12\frac{\PM_\PH^2}{\sin^2\theta_\Pw}\frac{k_2^\mu}{k_2^2}
			\frac{k_3^\nu}{k_3^2}\frac{k_4^\sigma}{k_4^2}
\right)
\cr
% 74
\begin{picture}(50,50)(0,0)
\Line(25,25)(0,0)\Text(12,13)[rb]{$\scriptscriptstyle \phi_\PZ(k_1)$}
\Line(25,25)(0,50)\Text(12,37)[rt]{$\scriptscriptstyle \PZ(k_2)^\mu$}
\ArrowLine(25,25)(50,0)\Text(38,13)[lb]{$\scriptscriptstyle \phi_\PW(k_3)$}
\ArrowLine(50,50)(25,25)\Text(38,37)[lt]{$\scriptscriptstyle \PW(k_4)^\nu$}
\end{picture}
&
-\Pg_\Pe^2\left(\frac1{2\cos\theta_\Pw}g^{\mu\nu}+\frac{\PM_\PH^2}{4\sin^2\theta_\Pw\cos\theta_\Pw}\frac{k_2^\mu}{k_2^2}\frac{k_4^\nu}{k_4^2}\right)
\cr
% 75 zie 74
% 76
\begin{picture}(50,50)(0,0)
\Line(25,25)(0,0)\Text(12,13)[rb]{$\scriptscriptstyle \phi_\PZ(k_1)$}
\Line(25,25)(0,50)\Text(12,37)[rt]{$\scriptscriptstyle \PZ(k_2)^\mu$}
\ArrowLine(50,0)(25,25)\Text(38,13)[lb]{$\scriptscriptstyle \phi_\PW(k_3)$}
\ArrowLine(25,25)(50,50)\Text(38,37)[lt]{$\scriptscriptstyle \phi_\PW(k_4)$}
\end{picture}
&
\frac14\Pg_\Pz^2\frac{\PM_\PH^2}{\PM_\PZ}\frac{k_2^\mu}{k_2^2}
\cr
% 77
\begin{picture}(50,50)(0,0)
\Line(25,25)(0,0)\Text(12,13)[rb]{$\scriptscriptstyle \phi_\PZ(k_1)$}
\Line(25,25)(0,50)\Text(12,37)[rt]{$\scriptscriptstyle \PZ(k_2)^\mu$}
\Line(50,0)(25,25)\Text(38,13)[lb]{$\scriptscriptstyle \PZ(k_3)^\nu$}
\Line(25,25)(50,50)\Text(38,37)[lt]{$\scriptscriptstyle \PZ(k_4)^\sigma$}
\end{picture}
&
-\frac12\Pg_\Pz^2\PM_\PZ\left(
	g^{\nu\sigma}\frac{k_2^\mu}{k_2^2}
	+g^{\mu\sigma}\frac{k_3^\nu}{k_3^2}
	+g^{\mu\nu}\frac{k_4^\sigma}{k_4^2}
	+\frac32\PM_\PH^2\frac{k_2^\mu}{k_2^2}\frac{k_3^\nu}{k_3^2}
																		\frac{k_4^\sigma}{k_4^2}
\right)
\cr
}

\subsection{Quadruple boson couplings with multiple $\phi_\PZ$ and no $\PH$}
\fmruletab{
% 78
\begin{picture}(50,50)(0,0)
\Line(25,25)(0,0)\Text(12,13)[rb]{$\scriptscriptstyle \phi_\PZ(k_1)$}
\Line(25,25)(0,50)\Text(12,37)[rt]{$\scriptscriptstyle \phi_\PZ(k_2)$}
\ArrowLine(50,0)(25,25)\Text(38,13)[lb]{$\scriptscriptstyle \PW(k_3)^\mu$}
\ArrowLine(25,25)(50,50)\Text(38,37)[lt]{$\scriptscriptstyle \PW(k_4)^\nu$}
\end{picture}
&
i\Pg_\Pw^2\left(\frac12g^{\mu\nu}-\frac14\PM_\PH^2\frac{k_3^\mu}{k_3^2}
																\frac{k_4^\nu}{k_4^2}\right)
\cr
% 79
\begin{picture}(50,50)(0,0)
\Line(25,25)(0,0)\Text(12,13)[rb]{$\scriptscriptstyle \phi_\PZ(k_1)$}
\Line(25,25)(0,50)\Text(12,37)[rt]{$\scriptscriptstyle \phi_\PZ(k_2)$}
\ArrowLine(25,25)(50,0)\Text(38,13)[lb]{$\scriptscriptstyle \phi_\PW(k_3)$}
\ArrowLine(50,50)(25,25)\Text(38,37)[lt]{$\scriptscriptstyle \PW(k_4)^\mu$}
\end{picture}
&
\frac i4\Pg_\Pw^2\frac{\PM_\PH^2}{\PM_\PW}\frac{k_4^\mu}{k_4^2}
\cr
% 80 zie 79
% 81
\begin{picture}(50,50)(0,0)
\Line(25,25)(0,0)\Text(12,13)[rb]{$\scriptscriptstyle \phi_\PZ(k_1)$}
\Line(25,25)(0,50)\Text(12,37)[rt]{$\scriptscriptstyle \phi_\PZ(k_2)$}
\ArrowLine(50,0)(25,25)\Text(38,13)[lb]{$\scriptscriptstyle \phi_\PW(k_3)$}
\ArrowLine(25,25)(50,50)\Text(38,37)[lt]{$\scriptscriptstyle \phi_\PW(k_4)$}
\end{picture}
&
-\frac i4\Pg_\Pw^2\frac{\PM_\PH^2}{\PM_\PW^2}
\cr
% 82
\begin{picture}(50,50)(0,0)
\Line(25,25)(0,0)\Text(12,13)[rb]{$\scriptscriptstyle \phi_\PZ(k_1)$}
\Line(25,25)(0,50)\Text(12,37)[rt]{$\scriptscriptstyle \phi_\PZ(k_2)$}
\Line(50,0)(25,25)\Text(38,13)[lb]{$\scriptscriptstyle \PZ(k_3)^\mu$}
\Line(25,25)(50,50)\Text(38,37)[lt]{$\scriptscriptstyle \PZ(k_4)^\nu$}
\end{picture}
&
\frac i2\Pg_\Pz^2\left(
	g^{\mu\nu}
	+\frac32\PM_\PH^2\frac{k_3^\mu}{k_3^2}\frac{k_4^\nu}{k_4^2}
\right)
\cr
% 83
\begin{picture}(50,50)(0,0)
\Line(25,25)(0,0)\Text(12,13)[rb]{$\scriptscriptstyle \phi_\PZ(k_1)$}
\Line(25,25)(0,50)\Text(12,37)[rt]{$\scriptscriptstyle \phi_\PZ(k_2)$}
\Line(50,0)(25,25)\Text(38,13)[lb]{$\scriptscriptstyle \phi_\PZ(k_3)$}
\Line(50,50)(25,25)\Text(38,37)[lt]{$\scriptscriptstyle \PZ(k_4)^\mu$}
\end{picture}
&
\frac34\Pg_\Pz^2\frac{\PM_\PH^2}{\PM_\PZ}\frac{k_4^\mu}{k_4^2}
\cr
% 84
\begin{picture}(50,50)(0,0)
\Line(25,25)(0,0)\Text(12,13)[rb]{$\scriptscriptstyle \phi_\PZ(k_1)$}
\Line(25,25)(0,50)\Text(12,37)[rt]{$\scriptscriptstyle \phi_\PZ(k_2)$}
\Line(50,0)(25,25)\Text(38,13)[lb]{$\scriptscriptstyle \phi_\PZ(k_3)$}
\Line(50,50)(25,25)\Text(38,37)[lt]{$\scriptscriptstyle \phi_\PZ(k_4)$}
\end{picture}
&
-\frac34i\Pg_\Pz^2\frac{\PM_\PH^2}{\PM_\PZ^2}
\cr
}
\subsection{Quadruple boson couplings with one $\PH$}
\fmruletab{
% 85
\begin{picture}(50,50)(0,0)
\Line(25,25)(0,0)\Text(12,13)[rb]{$\scriptscriptstyle \PH(k_1)$}
\ArrowLine(0,50)(25,25)\Text(12,37)[rt]{$\scriptscriptstyle \PW(k_2)^\mu$}
\ArrowLine(25,25)(50,0)\Text(38,13)[lb]{$\scriptscriptstyle \PW(k_3)^\nu$}
\Line(50,50)(25,25)\Text(38,37)[lt]{$\scriptscriptstyle \PB(k_4)^\sigma$}
\end{picture}
&
-\frac i2\Pg_\Pe\Pg_\Pw\PM_\PW\left(
	g^{\nu\sigma}\frac{k_2^\mu}{k_2^2}
	+g^{\mu\sigma}\frac{k_3^\nu}{k_3^2}
\right)
\cr
% 86
\begin{picture}(50,50)(0,0)
\Line(25,25)(0,0)\Text(12,13)[rb]{$\scriptscriptstyle \PH(k_1)$}
\ArrowLine(25,25)(0,50)\Text(12,37)[rt]{$\scriptscriptstyle \phi_\PW(k_2)$}
\ArrowLine(50,0)(25,25)\Text(38,13)[lb]{$\scriptscriptstyle \PW(k_3)^\mu$}
\Line(50,50)(25,25)\Text(38,37)[lt]{$\scriptscriptstyle \PB(k_4)^\nu$}
\end{picture}
&
\frac i2\Pg_\Pe\Pg_\Pw g^{\mu\nu}
\cr
% 87 zie 86
% 88
\begin{picture}(50,50)(0,0)
\Line(25,25)(0,0)\Text(12,13)[rb]{$\scriptscriptstyle \PH(k_1)$}
\Line(25,25)(0,50)\Text(12,37)[rt]{$\scriptscriptstyle \PZ(k_2)^\mu$}
\ArrowLine(50,0)(25,25)\Text(38,13)[lb]{$\scriptscriptstyle \PW(k_3)^\nu$}
\ArrowLine(25,25)(50,50)\Text(38,37)[lt]{$\scriptscriptstyle \PW(k_4)^\sigma$}
\end{picture}
&
\frac i2\Pg_\Pe^2\PM_\PZ\left(
	g^{\mu\sigma}\frac{k_3^\nu}{k_3^2}
	+g^{\mu\nu}\frac{k_4^\sigma}{k_4^2}
\right)
\cr
% 89
\begin{picture}(50,50)(0,0)
\Line(25,25)(0,0)\Text(12,13)[rb]{$\scriptscriptstyle \PH(k_1)$}
\Line(25,25)(0,50)\Text(12,37)[rt]{$\scriptscriptstyle \PZ(k_2)^\mu$}
\ArrowLine(25,25)(50,0)\Text(38,13)[lb]{$\scriptscriptstyle \phi_\PW(k_3)$}
\ArrowLine(50,50)(25,25)\Text(38,37)[lt]{$\scriptscriptstyle \PW(k_4)^\nu$}
\end{picture}
&
-\frac i2\frac{\Pg_\Pe^2}{\cos\theta_\Pw}g^{\mu\nu}
\cr
% 90 zie 89
}

\subsection{Quadruple boson couplings with multiple $\PH$}
\fmruletab{
% 91
\begin{picture}(50,50)(0,0)
\Line(25,25)(0,0)\Text(12,13)[rb]{$\scriptscriptstyle \PH(k_1)$}
\Line(25,25)(0,50)\Text(12,37)[rt]{$\scriptscriptstyle \PH(k_2)$}
\ArrowLine(50,0)(25,25)\Text(38,13)[lb]{$\scriptscriptstyle \PW(k_3)^\mu$}
\ArrowLine(25,25)(50,50)\Text(38,37)[lt]{$\scriptscriptstyle \PW(k_4)^\nu$}
\end{picture}
&
i\Pg_\Pw^2\left(\frac12g^{\mu\nu}
	-\frac14\PM_\PH^2\frac{k_3^\mu}{k_3^2}\frac{k_4^\nu}{k_4^2}\right)
\cr
% 92
\begin{picture}(50,50)(0,0)
\Line(25,25)(0,0)\Text(12,13)[rb]{$\scriptscriptstyle \PH(k_1)$}
\Line(25,25)(0,50)\Text(12,37)[rt]{$\scriptscriptstyle \PH(k_2)$}
\ArrowLine(25,25)(50,0)\Text(38,13)[lb]{$\scriptscriptstyle \phi_\PW(k_3)$}
\ArrowLine(50,50)(25,25)\Text(38,37)[lt]{$\scriptscriptstyle \PW(k_4)^\mu$}
\end{picture}
&
\frac i4\Pg_\Pw^2\frac{\PM_\PH^2}{\PM_\PW}\frac{k_4^\mu}{k_4^2}
\cr
% 93 zie 92
% 94
\begin{picture}(50,50)(0,0)
\Line(25,25)(0,0)\Text(12,13)[rb]{$\scriptscriptstyle \PH(k_1)$}
\Line(25,25)(0,50)\Text(12,37)[rt]{$\scriptscriptstyle \PH(k_2)$}
\ArrowLine(50,0)(25,25)\Text(38,13)[lb]{$\scriptscriptstyle \phi_\PW(k_3)$}
\ArrowLine(25,25)(50,50)\Text(38,37)[lt]{$\scriptscriptstyle \phi_\PW(k_4)$}
\end{picture}
&
-\frac i4\Pg_\Pw^2\frac{\PM_\PH^2}{\PM_\PW^2}
\cr
% 95
\begin{picture}(50,50)(0,0)
\Line(25,25)(0,0)\Text(12,13)[rb]{$\scriptscriptstyle \PH(k_1)$}
\Line(25,25)(0,50)\Text(12,37)[rt]{$\scriptscriptstyle \PH(k_2)$}
\Line(25,25)(50,0)\Text(38,13)[lb]{$\scriptscriptstyle \PZ(k_3)^\mu$}
\Line(50,50)(25,25)\Text(38,37)[lt]{$\scriptscriptstyle \PZ(k_4)^\nu$}
\end{picture}
&
\frac i2\Pg_\Pz^2\left(
	g^{\mu\nu}
	+\frac12\PM_\PH^2\frac{k_3^\mu}{k_3^2}\frac{k_4^\nu}{k_4^2}
\right)
\cr
% 96
\begin{picture}(50,50)(0,0)
\Line(25,25)(0,0)\Text(12,13)[rb]{$\scriptscriptstyle \PH(k_1)$}
\Line(25,25)(0,50)\Text(12,37)[rt]{$\scriptscriptstyle \PH(k_2)$}
\Line(50,0)(25,25)\Text(38,13)[lb]{$\scriptscriptstyle \phi_\PZ(k_3)$}
\Line(25,25)(50,50)\Text(38,37)[lt]{$\scriptscriptstyle \PZ(k_4)^\mu$}
\end{picture}
&
\frac14\Pg_\Pz^2\frac{\PM_\PH^2}{\PM_\PZ}\frac{k_4^\mu}{k_4^2}
\cr
% 97
\begin{picture}(50,50)(0,0)
\Line(25,25)(0,0)\Text(12,13)[rb]{$\scriptscriptstyle \PH(k_1)$}
\Line(25,25)(0,50)\Text(12,37)[rt]{$\scriptscriptstyle \PH(k_2)$}
\Line(50,0)(25,25)\Text(38,13)[lb]{$\scriptscriptstyle \phi_\PZ(k_3)$}
\Line(25,25)(50,50)\Text(38,37)[lt]{$\scriptscriptstyle \phi_\PZ(k_4)$}
\end{picture}
&
-\frac i4\Pg_\Pw^2\frac{\PM_\PH^2}{\PM_\PW^2}
\cr
% 98
\begin{picture}(50,50)(0,0)
\Line(25,25)(0,0)\Text(12,13)[rb]{$\scriptscriptstyle \PH(k_1)$}
\Line(25,25)(0,50)\Text(12,37)[rt]{$\scriptscriptstyle \PH(k_2)$}
\Line(50,0)(25,25)\Text(38,13)[lb]{$\scriptscriptstyle \PH(k_3)$}
\Line(25,25)(50,50)\Text(38,37)[lt]{$\scriptscriptstyle \PH(k_4)$}
\end{picture}
&
-\frac34i\Pg_\Pw^2\frac{\PM_\PH^2}{\PM_\PW^2}
\cr
}

%\section{(Un)physical Particles}
%The $\phi_W$ and $\phi_Z$ fields are unphysical. This means that they
%cannot be external lines in a Feynman graph. The pole at $k^2=0$ that
%occurs in their propagators is canceled by the poles in the interaction
%vertices that the $W$ and $Z$ particles have. The consequence is that
%these particles cannot travel over macroscopic distances. 

\end{document}